\documentclass[3p,number,sort&compress,twocolumn]{elsarticle}

\usepackage{stfloats}

\usepackage[dvipsnames]{xcolor}

\usepackage{amsmath}
\usepackage{lineno, hyperref}

\journal{Nuc. Instr. and Meth. A}

\biboptions{compress}

\begin{document}

\begin{frontmatter}

\title{A high-finesse Fabry-Perot cavity with a frequency-doubled green laser for precision Compton polarimetry at Jefferson Lab}





\author[1,2]{A.~Rakhman\corref{cor1}}
\ead{rahim@ornl.gov}
\cortext[cor1]{Corresponding author: Research Accelerator Division, Spallation Neutron Source, Oak Ridge National Laboratory, Oak Ridge, TN 37831, USA. Tel.: +1 315 391 4622; Fax.: +1 865 576 9209.}

\author[3]{M.~Hafez}
\author[4]{S.~Nanda}
\author[5,6]{F.~Benmokhtar}
\author[4]{A.~Camsonne}
\author[7]{G.D.~Cates}
\author[4,7]{M.M.~Dalton}
\author[5]{G.B.~Franklin}
\author[5,8]{M.~Friend}
\author[4]{R.W.~Michaels}
\author[7]{V.~Nelyubin}
\author[5,9]{D.S.~Parno}
\author[7]{K.D.~Paschke}
\author[5]{B.P.~Quinn}
\author[1]{P.A.~Souder}
\author[7]{W.A.~Tobias}

\address[1]{Syracuse University, Department of Physics, Syracuse, NY 13244, USA\fnref{label3}}

\address[2]{Research Accelerator Division, Spallation Neutron Source, Oak Ridge National Laboratory, Oak Ridge, TN 37831, USA}

\address[3]{Old Dominion University, Applied Research Center, Norfolk, VA 23529, USA}

\address[4]{Thomas Jefferson National Accelerator Facility, Newport News, VA 23606, USA}

\address[5]{Carnegie Mellon University, Department of Physics, Pittsburgh, PA 15213, USA}

\address[6]{Duquesne University, Pittsburgh, PA 15282, USA}

\address[7]{University of Virginia, Department of Physics, Charlottesville, VA 22904, USA}

\address[8]{High Energy Accelerator Research Organization (KEK), Tsukuba, Ibaraki, Japan}

\address[9]{University of Washington, Center for Experimental Nuclear Physics and Astrophysics and Department of Physics, Seattle, WA 98195, USA}

\begin{abstract}

A high-finesse Fabry-Perot cavity with a frequency-doubled continuous wave green laser (532~nm) has been built and installed in Hall A of Jefferson Lab for high precision Compton polarimetry. The infrared (1064~nm) beam from a ytterbium-doped fiber amplifier seeded by a Nd:YAG nonplanar ring oscillator laser is frequency doubled in a single-pass periodically poled MgO:LiNbO$_{3}$ crystal. The maximum achieved green power at 5 W infrared pump power is 1.74 W with a total conversion efficiency of 34.8\%. The green beam is injected into the optical resonant cavity and enhanced up to 3.7~kW with a corresponding enhancement of 3800. The polarization transfer function has been measured in order to determine the intra-cavity circular laser polarization within a measurement uncertainty of 0.7\%. The PREx experiment at Jefferson Lab used this system for the first time and achieved 1.0\% precision in polarization measurements of an electron beam with energy and current of 1.06~GeV and 50~$\mu$A. \\

\end{abstract}


\end{frontmatter}

%
\section{Introduction}\label{sec:intro}
\indent The Continuous Electron Beam Accelerator Facility (CEBAF) at Jefferson Lab (JLab) delivers a highly polarized electron beam in the 1 -- 11~GeV energy range for currents up to $\sim$ 200~$\mu$A. A key part of the JLab physics program consists of high-precision parity violating electron scattering experiments \cite{PRExPRL, QWeakPRL, PVDISNature} that require a fast and precise electron beam polarization measurement. Conventional polarimetry techniques such as M\o ller  \cite {MollerPaper} and Mott \cite {MottPaper} are destructive for the beam properties due to solid targets they use and cannot be operated simultaneously with the physics experiments.  They can only be operated at low intensity ($<$ 10~$\mu$A) and at low energy ($<$ 5~MeV) respectively; therefore, the physics experiments have to assume that beam polarization remains constant when the intensity or the energy of the beam is several orders of magnitude higher. Alternatively Compton polarimetry has been used with great success at JLab \cite{Baylac, HallCCompton} and elsewhere \cite{NIKHEF, HERA, LEP, SLDCompton, ELSA} to measure the electron beam polarization in a continuous and non-destructive manner. However, unlike at high-current ($>$ 20~mA) storage rings \cite{NIKHEF, HERA, ELSA} and at high-energy ($>$ 50~GeV) colliders \cite{LEP, SLDCompton}, the relatively low-current and low-energy nature of the CEBAF electron beam has required the use of high-finesse optical cavities \cite{JPJorda, NFalletto} to enhance the laser power beyond commercially available lasers to provide sufficient electron-photon collision luminosity.\\
\indent Compton polarimetry determines the electron beam polarization based on a well-known spin-dependent scattering cross-section \cite{LippsPhysica} between polarized electrons and photons. The longitudinal electron beam polarization $P_{\text{e}}$ is extracted from the experimental Compton scattering asymmetry $A_{\text{exp}}$, according to,
\begin{equation}\label{eq:CmptAsym}
A_{\text{exp}} = P_{\text{e}} P_{\gamma} A_{\text{th}},
\end{equation}
where $A_{\text{th}}$ is the theoretical Compton scattering asymmetry that can be derived by combining the calculated spin-dependent Compton cross section with the appropriate experimental response function \cite{Prescott}, and $P_{\gamma}$ is the circular photon beam polarization. The scattering asymmetry $A_{\text{exp}}$ can be measured based on the detection of either the scattered photons or the scattered electrons and depends on the energies of the colliding electron and photon beams. For an electron beam of 1.0~GeV, the maximum scattering asymmetry for infrared (1064~nm) photons is 1.8\%, whereas for green (532~nm) photons, it is 3.5\%. For an electron beam of 6~GeV, these values are about 10\% and 18\%, respectively. Therefore, Compton polarimetry at low beam energies ($<$ 2~GeV) is difficult to perform.\\
\indent External optical cavities have frequently been employed to increase the laser power from a light source through a process of coherent addition. Recently optical cavities have been widely employed in many areas such as X-/$\gamma$-ray generation via inverse-Compton scattering \cite{ICS-Huang, ICS-Graves}, high harmonic generation \cite{HHG} and electron beam polarimetry \cite{NFalletto, HERA_JIns}. According to the best of our knowledge, there are only three facilities in the world have ever built a Compton polarimeter with an optical cavity; Mainz Microtron (MAMI) \cite{MAMICompton}, Hadron Electron Ring Accelerator (HERA) \cite{HERA_JIns} and JLab \cite{NFalletto}. In Hall A of JLab, a Compton back-scattering polarimeter had been operational since 1999 \cite{Escoffier}. Its photon source had a typical circulating power of  $\sim$ 1.5~kW inside a monolithic Fabry-Perot cavity where the circularly polarized photons scattered from the longitudinally polarized electrons. It provided 1.0\% statistical (and 1.2\% systematic) precision with $\sim$ 25 minutes' data taking for 4.6~GeV, 40~$\mu$A electrons \cite{Escoffier}.  This polarimeter used a continuous-wave (CW) infrared (IR, 1064~nm) laser as its photon source and had serious limitations in providing a good signal-to-noise ratio for beam energies below 2~GeV \cite{ComptonCDR}. Under this circumstance, it is highly desirable to have a shorter-wavelength and higher-power light source since this will result in larger back-scattered electron and photon energies and a larger measured scattering asymmetry, and hence increased ability to control systematic uncertainties.\\
\indent The Lead Radius Experiment (PREx) \cite{PRExPRL} at JLab aims to provide the first model-independent evidence of the existence of a significant neutron skin in $^{208}$Pb. PREx proposed to obtain a statistical and systematic precisions of 3.0\% and 2.0\% respectively on the parity-violating electroweak asymmetry of polarized electrons from $^{208}$Pb to determine the neutron radius of $^{208}$Pb at 1.0\% accuracy. The uncertainty goal for beam polarimetry at 1.06~GeV was 1.0\%. In order to meet the high-precision requirement, the existing Compton polarimeter underwent an upgrade \cite{MeganNIMA, DianaNIMA} in 2010. The upgrade includes a new photon detector with an upgraded integrating data acquisition (DAQ) system, a new electron detector, and a new Compton photon source that includes a green laser (532~nm) with a new high-finesse Fabry-Perot cavity \cite{RakhmanThesis}.\\
\indent This paper will describe the upgraded experimental setup in Hall A at JLab, with an emphasis on the frequency-doubled green laser system and optical cavity. A unique technique to extract the laser beam polarization inside a resonant Fabry-Perot cavity will be given and finally, the first electron beam polarization measurement results from the PREx experiment \cite{PRExPRL} at a beam energy and current of 1.06~GeV and 50~$\mu$A will also be presented. With an enhancement factor and intra-cavity power of 3,800 and 3.7 kW, we report the highest power Compton polarimetry facility operating at 532~nm as compared to existing facilities such as MAMI \cite{MAMICompton} and Hall C at JLab \cite{HallCCompton}.
\section{Overview of upgraded Compton polarimeter}\label{sec:polarimeter}
\begin{figure}[t!]
    \centerline{\includegraphics[width=7.8cm]{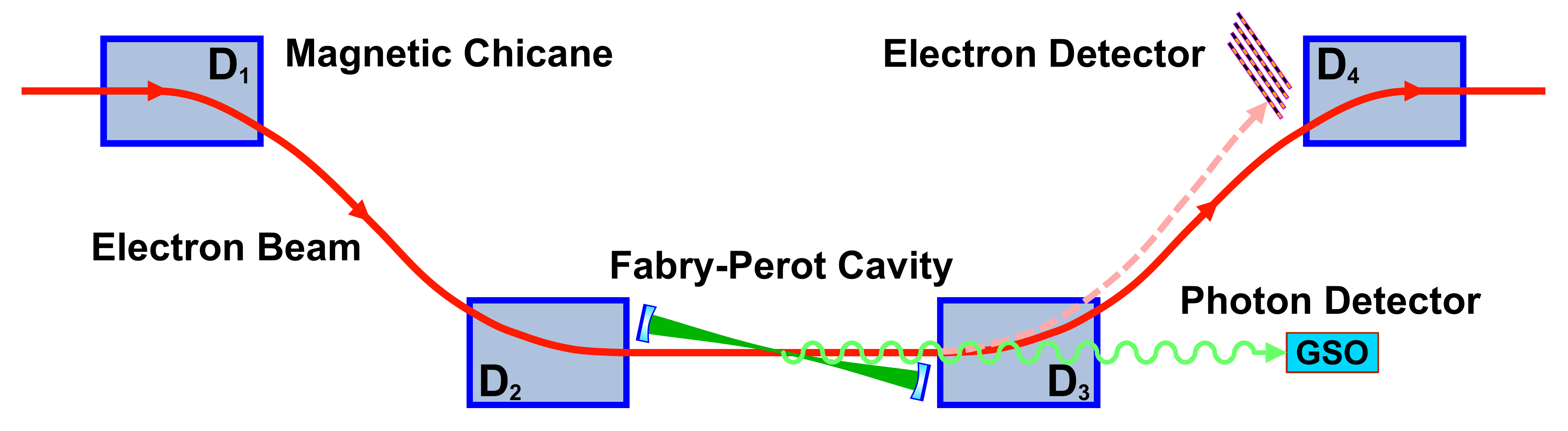}}
    \caption {A not-to-scale schematic of the upgraded Compton polarimeter in Hall A at JLab. A magnetic chicane consists of four identical magnetic dipoles (D$_{1}$, D$_{2}$, D$_{3}$ and D$_{4}$) that deflect the primary electron beam and send it through the Fabry-Perot cavity. The scattered Compton electrons and photons will be detected by electron and photon detectors, respectively.}
    \label{fig:Chicane}
\end{figure}
\begin{figure*}[b!]
\centering
    \centerline{\includegraphics[width=16.0cm]{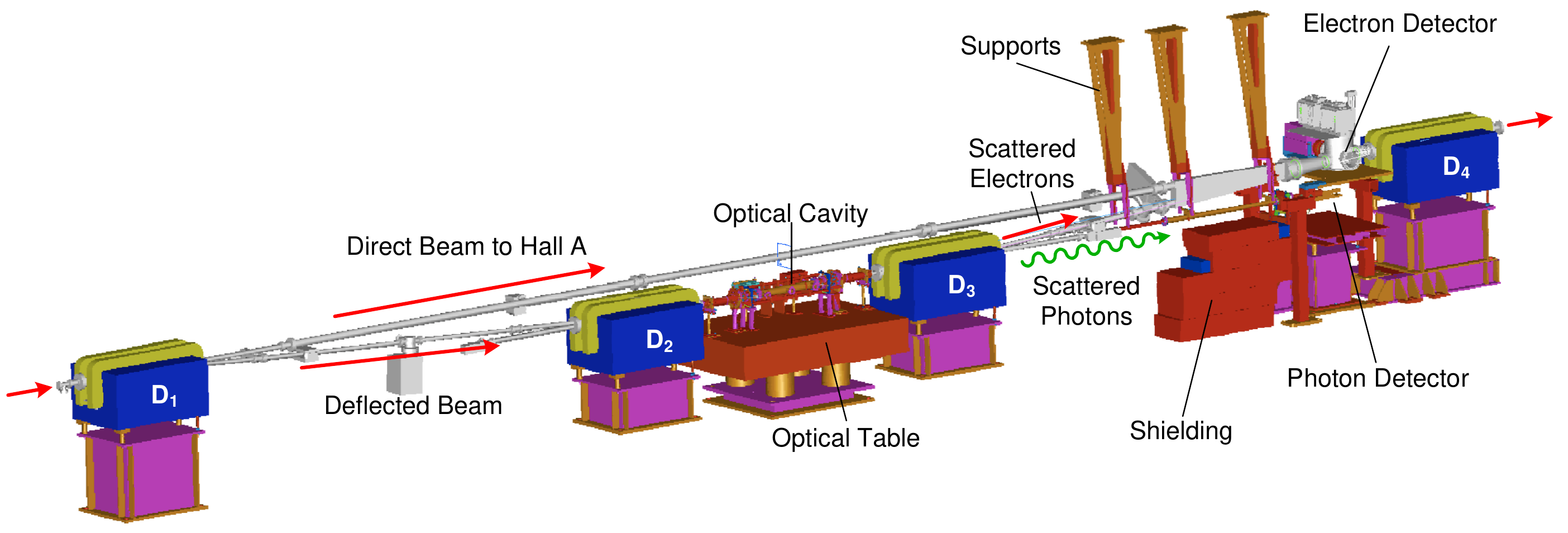}}
    \caption{A scale 3D view of the Compton polarimeter in the Hall A tunnel at JLab.}
    \label{fig:Chicane3D}
\end{figure*}
The upgraded Compton polarimeter in Hall A consists of a magnetic chicane, a photon source that is composed of a laser system and Fabry-Perot cavity, a photon detector, and an electron detector as shown in Fig. \ref{fig:Chicane}. The electron beam enters from the left. When the four identical dipoles of the chicane, referred to as D$_{1}$, D$_{2}$, D$_{3}$ and D$_{4}$, are energized, the beam is deflected vertically and crosses the photon beam enhanced in the Fabry-Perot cavity that is located at the center of the chicane. The Fabry-Perot cavity is enclosed in a vacuum chamber that is connected to electron beam pipes from both ends and sits on an optical table where the laser and other optical elements are located. The Compton Interaction Point (CIP) is at the center of the beam crossing in the cavity. The crossing angle between the electron and photon beams is 24~mrad. The electrons undergo Compton scattering with circularly polarized photons resonating in the Fabry-Perot cavity fed by a CW green laser (532~nm). The backscattered photons are detected in the photon detector while the scattered electrons can be detected in the electron detector located a few mm above the primary beam in front of D$_{4}$. Approximately one electron in every one billion undergoes Compton scattering. Unscattered electrons, separated from the scattered particles by D$_{3}$ in the chicane, continue on into the Hall A target for the primary experiment. Independent electron and photon analyses give complementary determinations of the electron beam polarization. A scale 3D view of the Compton polarimeter in the Hall A tunnel at JLab is shown in Fig. \ref{fig:Chicane3D}.\\
\indent The photon detector is located under dipole D$_{4}$ and consist of a single crystal of Ce-doped Gd$_{2}$SiO$_{5}$ (GSO) and a single PMT (Photomultiplier Tube). It is mounted on a motorized table with remote-controllable motion along both axes (horizontal and vertical) transverse to the electron beam direction. The scattered photons from the CIP go through a thin vacuum window and lead disk in the vacuum pipe before they hit the detector window. The photon detector DAQ uses a customized 12-bit Flash Analog-to-Digital Converter (FADC) that integrates the scintillation photon signals from the PMT and samples the data at rates of 200~mW \cite{MeganNIMA}.\\
\indent The electron detector consists of four silicon microstrip planes spaced horizontally 1 cm from each other with a 200 micron vertical rising offset between planes. Each plane consists of 192 strips of silicon with a width of 240 microns. The planes are inclined at an angle of 58~mrad from the vertical position. The detector is mounted on a translator with a remote-controllable stepper motor and can travel vertically up to 120~mm from the main beam.
\section{Optical and mechanical system}\label{sec:optics}
\begin{figure*}[b!]
\centering
    \centerline{\includegraphics[width=16.0cm]{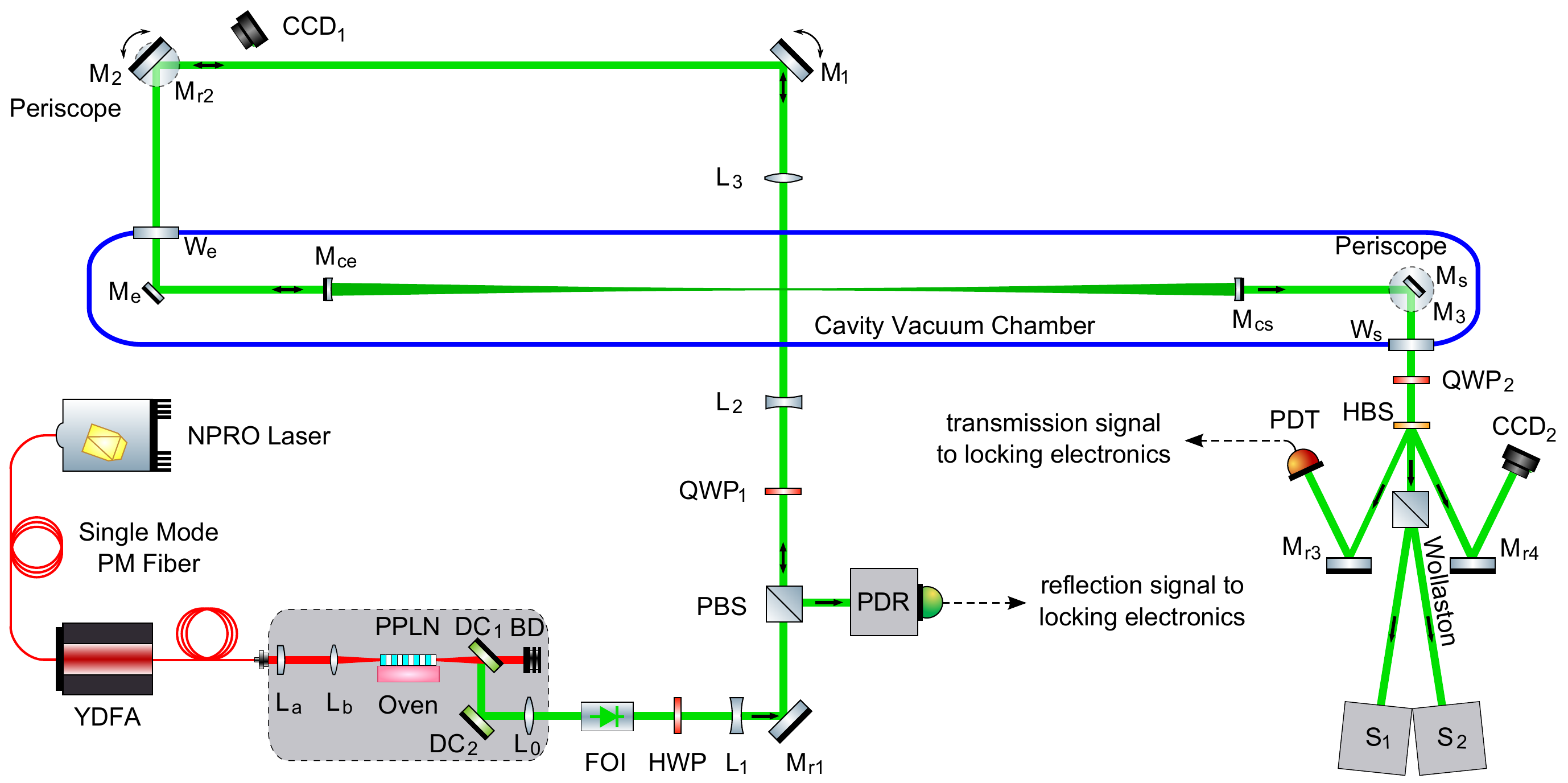}}
    \caption{The optical system consists of a NPRO seed laser, a fiber amplifier, a frequency doubling setup and a Fabry-Perot cavity enclosed in a vacuum chamber. (YDFA: ytterbium doped fiber amplifier; PPLN: periodically poled lithium niobate; DC$_{1}$, DC$_{2}$: dichroic mirrors; BD: beam dump; FOI: Faraday optical isolator; HWP: half-wave plate; PBS: polarizing beam splitter; QWP: quarter-wave plate; HBS: holographic beam sampler; M$_{ce}$, M$_{cs}$: concave cavity mirrors; CCD: cameras; PDR, PDT: photodiodes; L: lenses; M: plane mirrors; W: vacuum windows; S: photodetectors)}
    \label{fig:OpticalSetup}
\end{figure*}
The optical elements of the polarimeter are mounted on a vibration-damped optical table between dipoles D$_{2}$ and D$_{3}$ (Fig. \ref{fig:Chicane3D}). A schematic of the optical system of the polarimeter is illustrated in Fig. \ref{fig:OpticalSetup}. Based on their main functionality, we can categorize the optical elements into four groups necessary for achieving the required power and circular polarization in the cavity. The first group is the laser source that provides a green beam at the wavelength of 532~nm. The second group consists of the elements that transport, align and focus the incident beam for coupling to the optical cavity. The third group includes elements for controlling and measuring the polarization. The last group consists of elements which allow the use of the reflected beam from the cavity in the electronic feedback system to achieve frequency locking of the laser to the cavity.
\subsection{Frequency doubling}
\begin{figure}[t!]
    \centerline{\includegraphics[width=7.5cm]{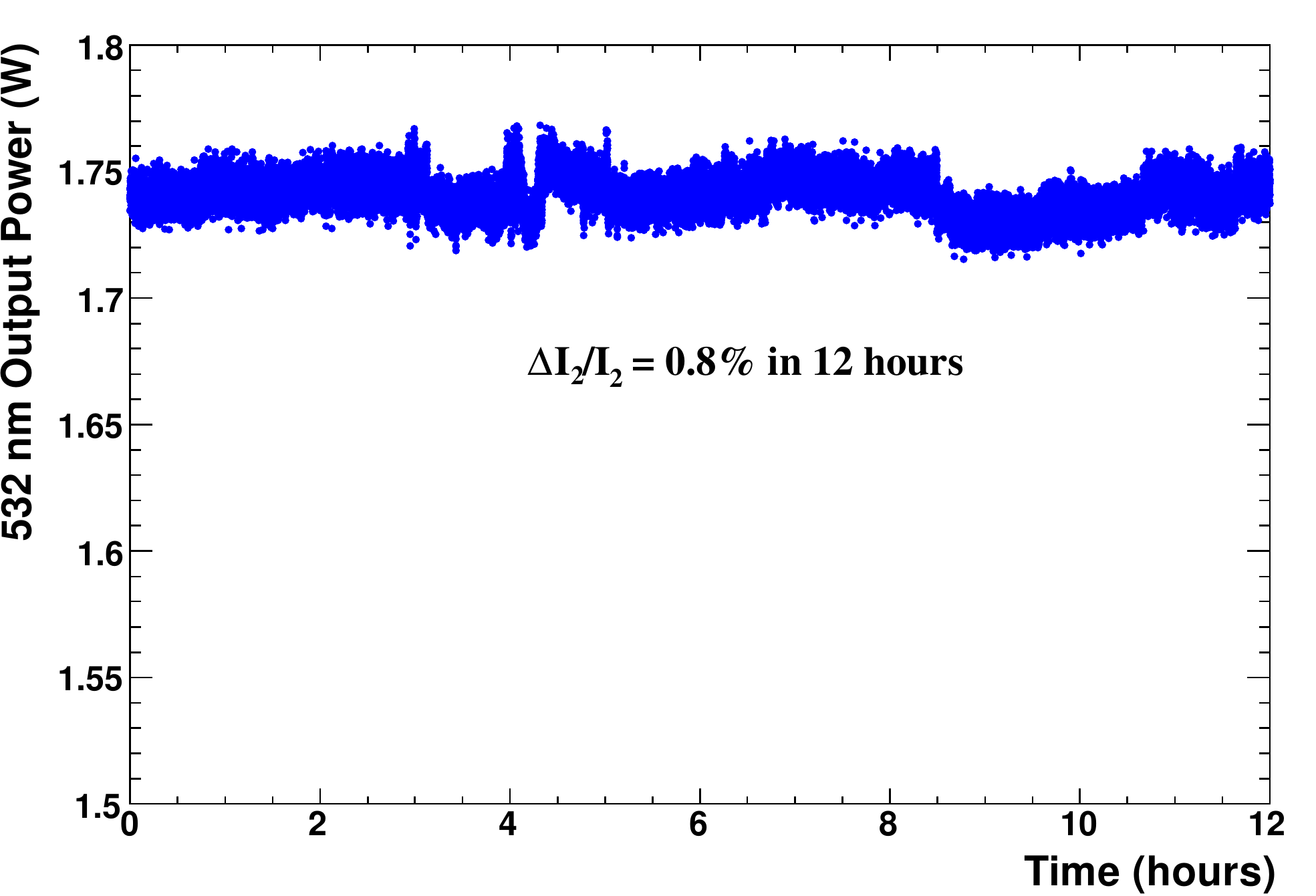}}
    \caption{Second harmonic power stability at the output of PPLN is monitored at 1.74~W for 12 hours.}
    \label{fig:PowerStability}
\end{figure}

\indent As shown in Fig. \ref{fig:OpticalSetup}, the seed laser is a narrow-linewidth ($<$ 5.0~kHz) diode pumped Nd:YAG nonplanar ring oscillator (NPRO) laser (Lightwave 126; JDSU) that delivers a linearly polarized CW IR (1064~nm) beam of up to 250~mW. The output of this laser has been fiber-coupled to a ytterbium-doped fiber amplifier (YDFA, YAR-10K-1064-LP-SF; IPG Photonics) through a single-mode polarization-maintaining (PM) fiber. The YDFA takes $\sim$ 10~mW of seed laser and provides a linearly polarized (extinction ratio: 20~dB) output with a maximum power of 10 W at 1064~nm in CW mode. The frequency doubling crystal is bulk periodically poled MgO:LiNbO$_{3}$ (PPLN, HC Photonics) doped with 5\% MgO in order to minimize photo-refractive damage. The crystal is 0.5~mm thick, 3~mm wide and 50~mm long, and the quasi-phase matching (QPM) period is 6.92~$\mu$m with a 50\% duty cycle. The input and output surfaces are antireflection (AR) coated for 1064~nm and 532~nm, respectively. The crystal is placed in an externally controlled, temperature-stabilized oven developed in-house. The oven is mounted on a stage composed of a four-axis tilt aligner (9071-V; Newport) in order to establish a precise alignment for phase matching. A temperature controller (TECSource 5305; Arroyo Instruments) with a nominal resolution of 0.01 $^{\circ}$C provides the required phase-matching temperature for second harmonic generation (SHG). The crystal optical axis is matched with the direction of the output polarization (vertical) of the YDFA. A pair of 0.5 inch lenses (L$_a$ and L$_b$ in Fig. \ref{fig:OpticalSetup}) with focal lengths of 13.8~mm and 15~mm (with a total effective focal length of $\sim$ 120~mm) are then used to focus the beam waist diameter ($\sim$ 80~$\mu$m) into the center of the crystal. This makes the Rayleigh length of the pump laser focus to be around 5~mm and ensures the beam size at the crystal entry and exit faces to be around 400~$\mu$m. This critical focusing also ensures the incoming beam is not clipped by the tiny crystal thickness while keeping the SHG efficiency close to the maximum. The generated green beam is separated from the residual IR beam after the crystal via a pair of 1.0 inch dichroic mirrors ($R >$ 99\% at 532~nm, $T >$ 95\% at 1064~nm)  noted as DC$_{1}$ and DC$_{2}$ (Altos Photonics). The lenses, oven and dichroic mirrors are all mounted on a separate linear translation stage seated on a rail, and the whole system is contained in an enclosure box.\\
\begin{figure}[t!]
    \centerline{\includegraphics[width=7.5cm]{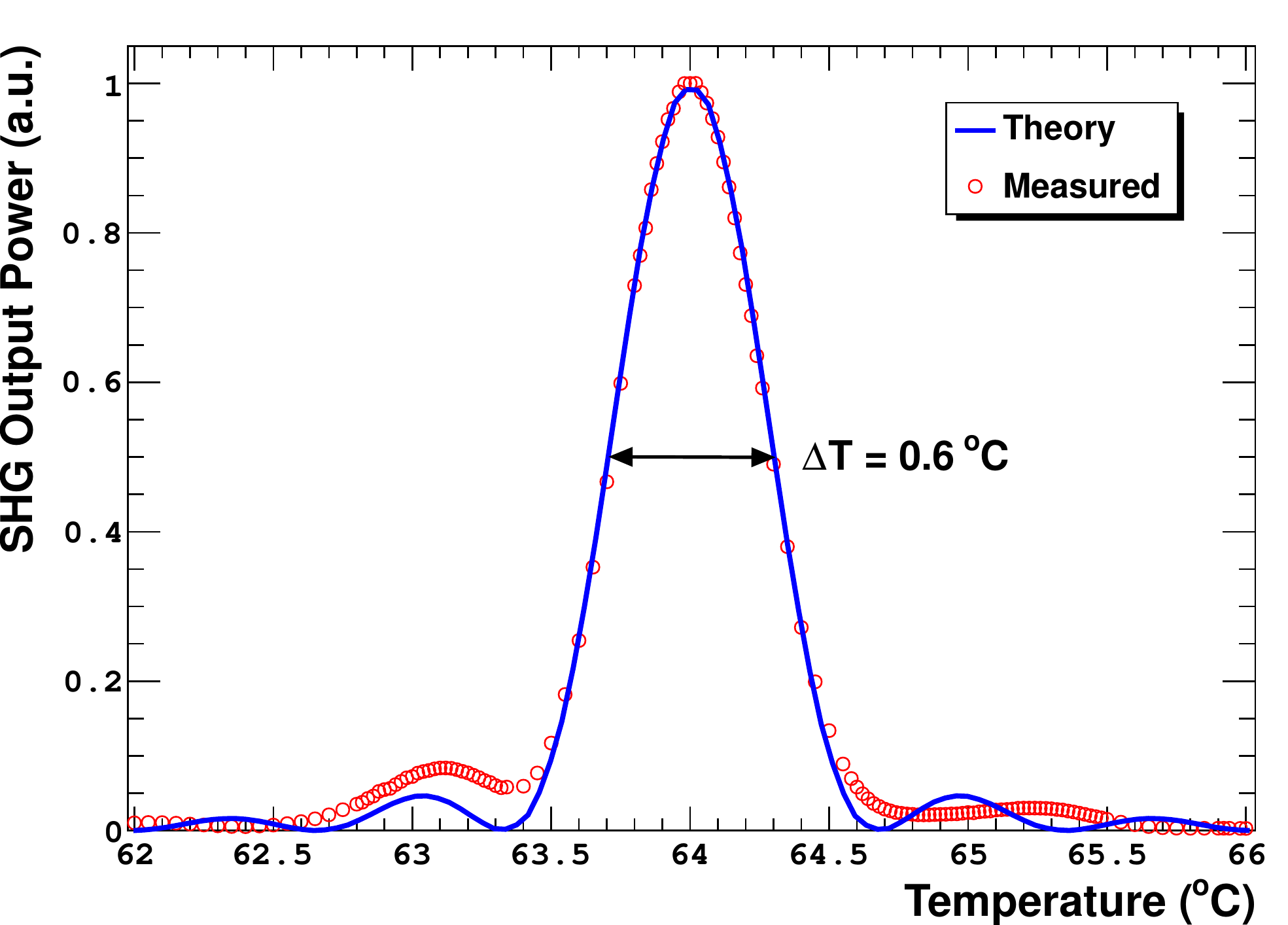}}
    \caption{Experimental (red circles) and theoretical \cite{Fejer, Jundt} (blue line) temperature tuning curve for PPLN crystal at beam waist diameter of 80~$\mu$m and IR pump power of 5 W.}\label{fig:TempScanFit}
\end{figure}
\indent With the experimental setup as described above, we have measured several properties, such as crystal temperature tuning curve, beam quality and power stability, of the second harmonic beam. The measured temperature tuning curve at 5~W IR pump power is shown in Fig. \ref{fig:TempScanFit}. The red circles show the experimental data while the blue line represents the theoretical fit for the first-order QPM interaction predicted by the Sellmeier equation for PPLN crystal \cite{Fejer, Jundt}. The experimental results give the full width at half-maximum (FWHM) phase-matching temperature bandwidth $\Delta T$ = 0.6~$^{\circ}$C at the phase-matching temperature 64.0~$^{\circ}$C.\\
\begin{figure}[t!]
    \centerline{\includegraphics[width=7.7cm]{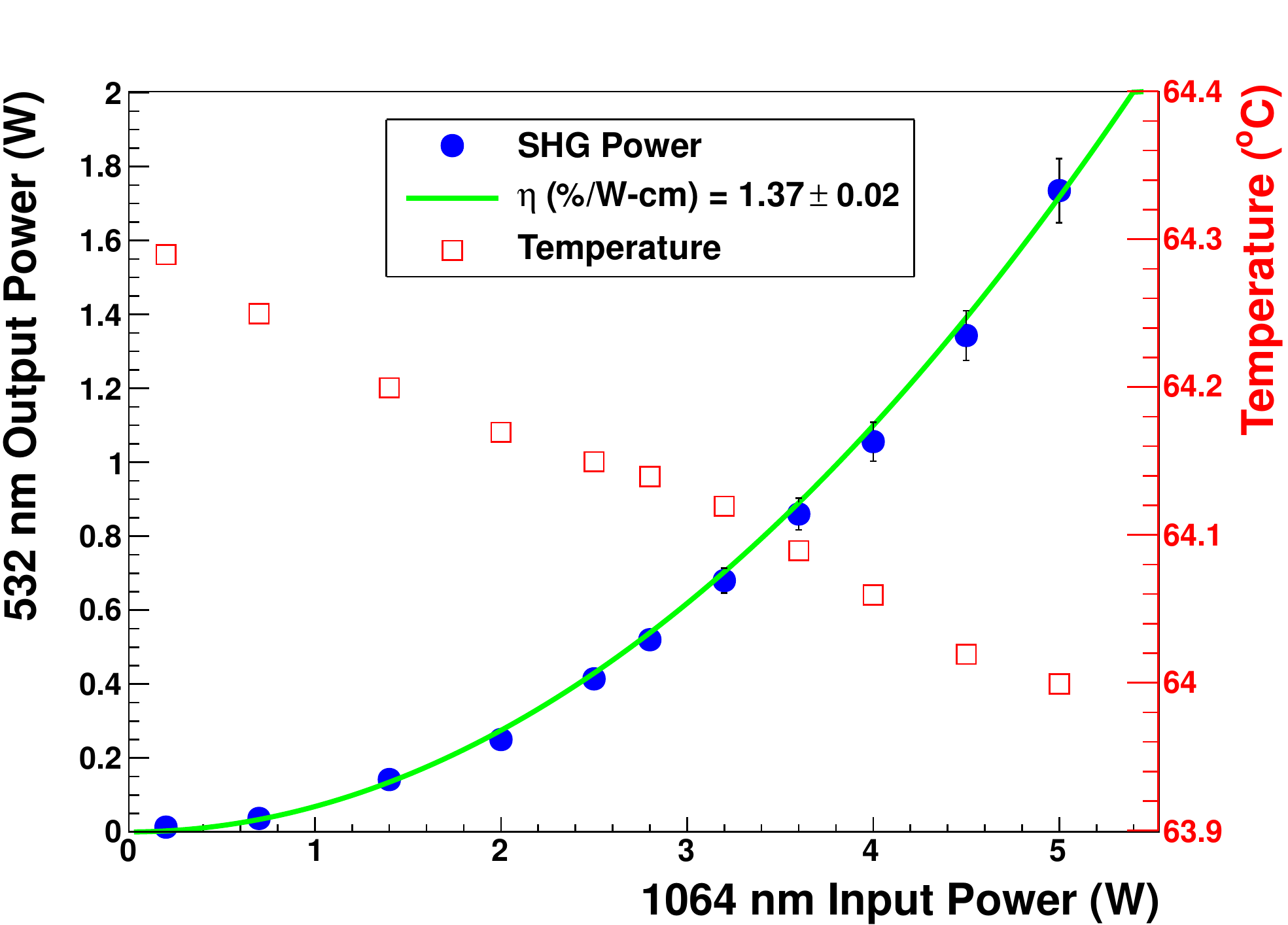}}
    \caption{Average green power (solid circles) output and corresponding phase matching temperature (open squares) at different IR pump power from the fiber amplifier. A theoretical fit (green line) defined by Eq. (\ref{eq:ConEff}) is used to extract the normalized SHG conversion efficiency.}
    \label{fig:PPLN_ConvEff}
\end{figure}
\indent Fig. \ref{fig:PPLN_ConvEff} shows the average green power (solid circles) output from PPLN and corresponding phase-matching temperature (open squares) at different IR pump powers from the fiber amplifier. It is important to note that the value of the phase-matching temperature changes slightly when the IR pump power is changed. The phase-matching temperature ranges between 64.0~$^{\circ}$C and 64.3~$^{\circ}$C for IR pump power between 200~mW and 5~W. At each power level the crystal is maintained at the peak phase-matching temperature to ensure maximum green power. For a loosely focused Gaussian beam, the normalized conversion efficiency $\eta_{nor}$ for first-order QPM is defined as \cite{DHum},
\begin{equation} \label{eq:ConEff}
I_{2} = \eta_{nor,QPM} L I_{1}^{2},
\end{equation}
\begin{equation}
\eta_{nor,QPM} = \frac{8\omega_{1}^{2}d^{2} L}{n_{1}^{2} n_{2} c^{3} \varepsilon_{0} \pi^{2}},
\end{equation}
where $I_{1}$ and $I_{2}$ are the fundamental and SHG powers, $\omega_{1}$ is the frequency of the fundamental beam, $d$ is the nonlinear coefficient of PPLN, $\varepsilon_{0}$ is the permittivity of free space, and $n_{1}$ and $n_{2}$ are the refractive indexes at these wavelengths. In Fig. \ref{fig:PPLN_ConvEff}, as the theoretical fit (green line) defined by Eq. (\ref{eq:ConEff}) shows, the green power varies quadratically with the IR pump power with a normalized conversion efficiency $\eta_{nor}$ of 1.37\%W$^{-1}$cm$^{-1}$. It is necessary to point out that all powers are direct measurements after DC$_{2}$ by an optical power meter (PM140; Thorlabs) without correcting for the residual reflection at the crystal face or losses in the dichroic mirrors. The experimental results show that there is no sign of saturation in the SHG power at 5~W IR pump power. This is mostly due to a larger beam waist size at the crystal center so that there is no pump depletion. Therefore, the achieved conversion efficiency is somewhat lower than the maximum conversion efficiency of 2.62\%W$^{-1}$cm$^{-1}$ provided by the vendor. All of these considerations suggest that thermal lens effects are negligible. The maximum achieved green power at 5~W IR pump power is 1.74~W with a total conversion efficiency of 34.8\%. A primary goal of this work is to develop a stable and good-quality green beam at the Watt level, so we did not pursue higher conversion efficiencies.\\
\begin{figure}[t!]
    \hspace{0.2cm}
    \centerline{\includegraphics[width=7.8cm]{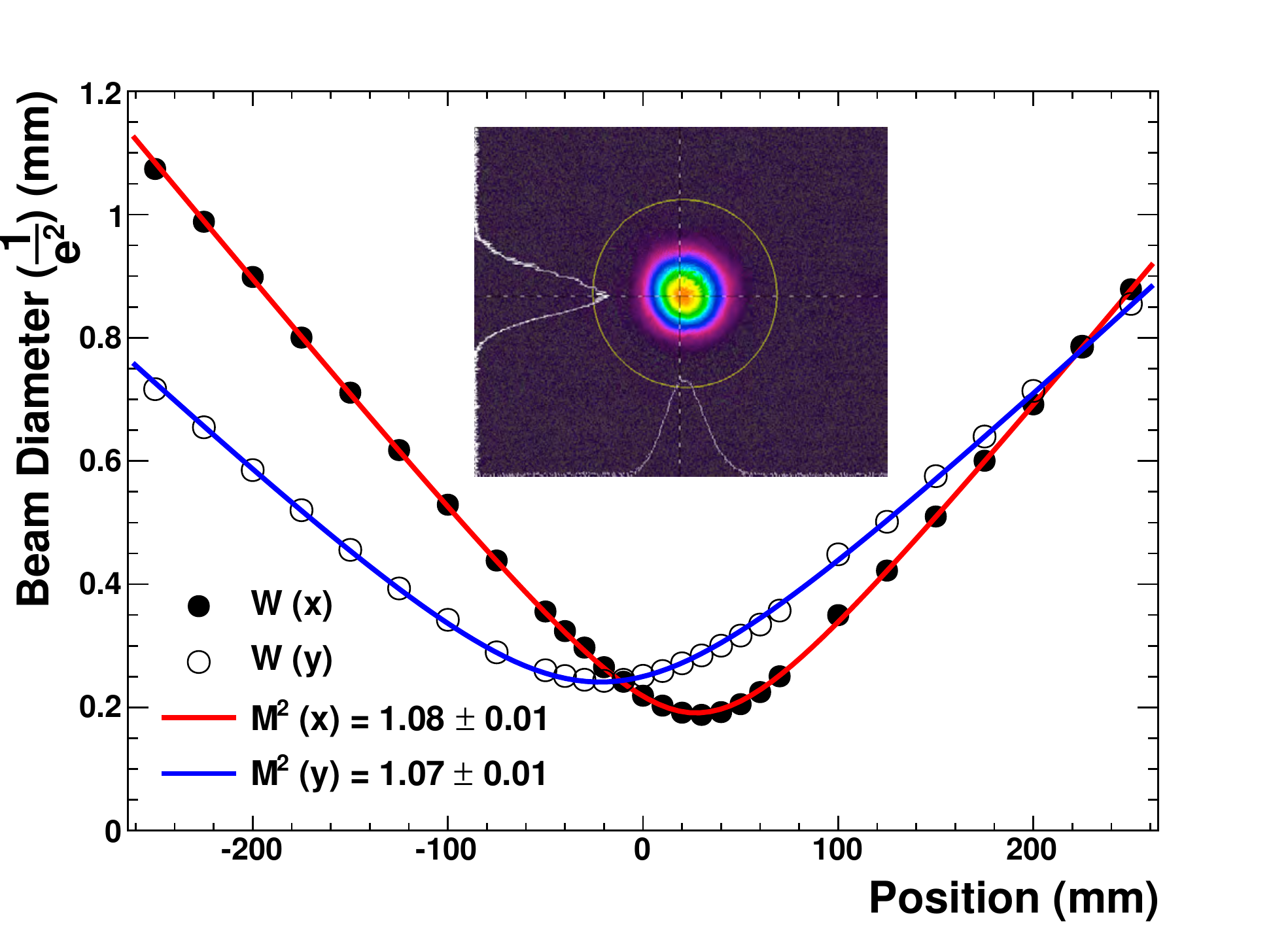}}
    \caption{The transverse intensity profile (shown in the inset) of the green beam measured by a laser beam profiler at 1.74~W green power. Closed and open circles are the beam diameters in the horizontal ($x$) and vertical ($y$) directions and continuous lines show the theoretical fits to extract the $M^2$ \cite{SiegmanBook} values in both directions.}
    \label{fig:PPLN_M2}
\end{figure}
\indent The quality of the green beam is of major importance in power enhancement with an optical cavity, as any deviation from a perfect Gaussian profile will lead to a rejection of power by the high-finesse cavity and will not contribute to the power buildup inside the Fabry-Perot cavity. In order to check the beam quality, we focused the green beam after the PPLN into a 200~$\mu$m waist diameter with an additional lens and measured the beam size at different axial positions. Fig. \ref{fig:PPLN_M2} shows the green beam profile at 1.74~W monitored by a laser beam profiler (LBA-FW-SCOR20; Spiricon) along with the theoretical fit. Closed and open circles are the measured beam diameters in the horizontal ($x$) and vertical ($y$) directions. The result of the theoretical fits (solid lines) to the experimental data determines an $M^2$ \cite{SiegmanBook} value of $\sim$ 1.1 in both dimensions confirming the TEM$_{00}$ spatial mode in the SHG process. However, there is a small astigmatism mainly arising from the difference in refractive indexes along the horizontal and vertical axes of the PPLN crystal. As shown in Fig. \ref{fig:PowerStability}, the peak-to-peak stability of green power at 1.74~W is 0.8\% for the entire period of 12 hours, while the fundamental IR power stability from the YDFA  is measured to be better than 0.6\% over 4 hours.\\
\indent The frequency-doubled green beam based on single-pass SHG in a PPLN crystal pumped by a YDFA has been used with the high-finesse Fabry-Perot cavity as a photon source for the Compton polarimeter during the three-month run period of the PREx experiment \cite{PRExPRL} in 2010. In the following we describe how we establish a frequency locking of this beam to a resonant frequency of the high-finesse Fabry-Perot cavity.
\subsection{High-finesse Fabry-Perot cavity}\label{sec:FPCavity}
\begin{figure}[t!]
    \vspace{-0.3cm}
    \hspace{0.2cm}
    \centerline{\includegraphics[width=8.2cm]{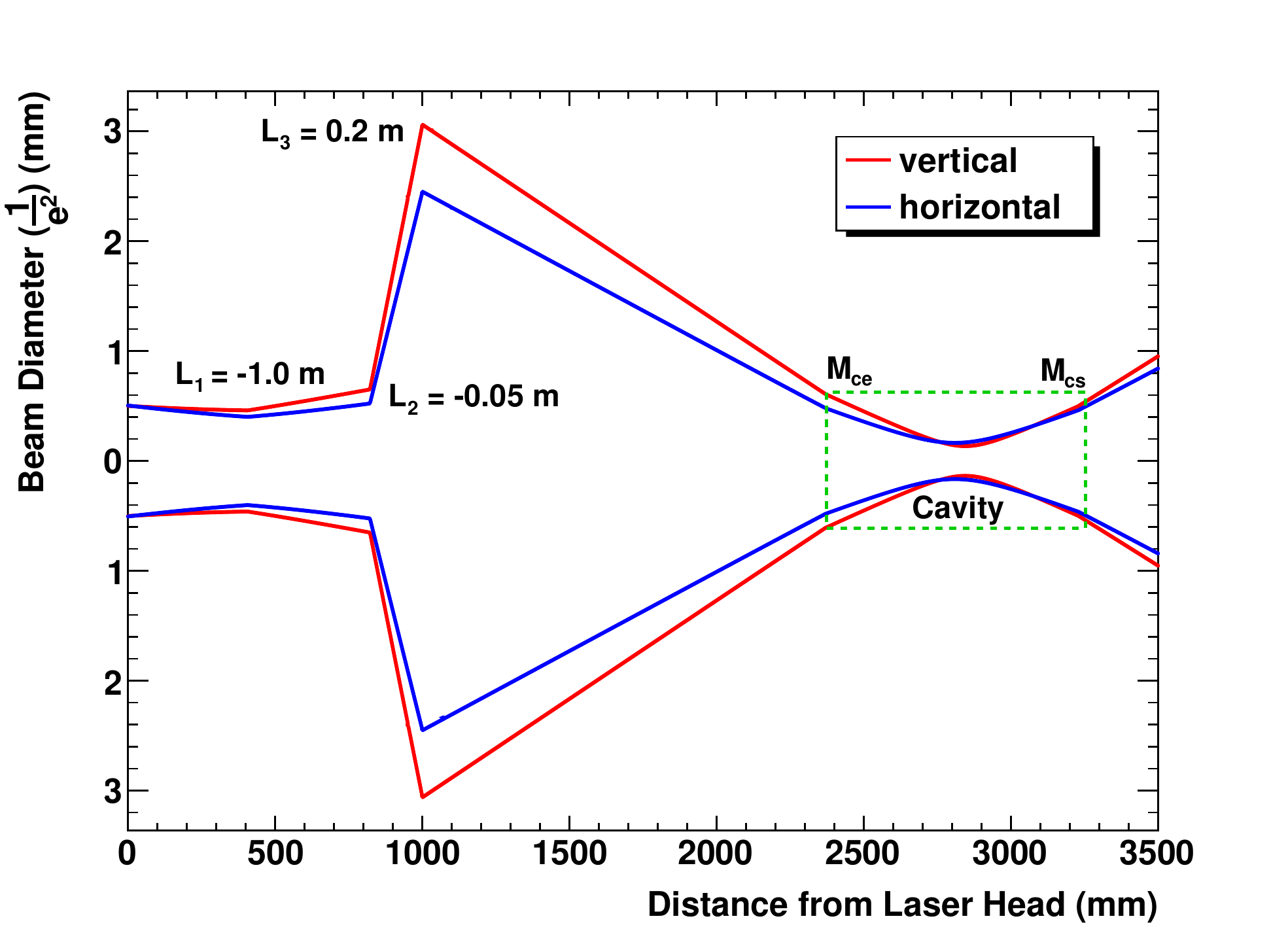}}
    \caption{The calculated beam size (beam envelope) versus the distance along the beam path from the frequency doubling setup.}\label{fig:BeamTransport}
\end{figure}

The  green beam is propagated through a series of optical elements (Fig. \ref{fig:OpticalSetup}) to properly match the fundamental cavity mode (TEM$_{00}$) at the injection point. After exiting the PPLN crystal and reflecting from two dichroic mirrors (DC$_1$ and DC$_2$), the green beam is focused by a lens L$_{0}$ ($f_0$ = 75~mm) at 6.0 cm from the frequency doubling setup with waist diameters 370~$\mu$m and 450~$\mu$m in the horizontal and vertical directions. The focusing of the incident beam at the CIP is accomplished by three lenses denoted L$_{1}$, L$_{2}$ and L$_{3}$, respectively. A diverging lens L$_1$ ($f_1$ = -1.0~m) at 405~mm from the frequency-doubling setup nearly collimates the beam after it passes through an optical isolator (isolation: 40~dB, IO-3-532-HP; Thorlabs) and a half-wave plate (HWP). In order to achieve the correct beam waist size at the cavity center, we have two more lenses, L$_2$ and L$_3$ ($f_2$ = -50~mm, $f_3$ = 200~mm), installed at 820~mm and 1.0~m from the frequency-doubling setup. Here in our setup, L$_1$ and L$_2$ are on fixed mounts and L$_3$ is mounted on a remote controlled translation stage (M-605.1DD; Physik Instrumente). This allows fine tuning and precise alignment of the laser beam to the cavity by using another pair of remotely controlled mirrors (M$_1$ and M$_2$) that allow four degrees of motion (2 translations, 2 rotations) for the laser beam with respect to the optical axis of the cavity. Both the proper focusing and precise alignment of the impinging laser beam to the cavity are needed in order to achieve good mode matching. The mirrors M$_{r1}$, M$_{r2}$, M$_{e}$ and M$_{s}$ are fixed at 45$^{\circ}$ with respect to the incident beam. A camera (CCD$_{1}$) facing the mirror M$_{r2}$ monitors the positions of the incident and reflected beams from the cavity. Another camera (CCD$_{2}$) at the cavity exit is used for monitoring the profile of the transmitted beam from the cavity.\\
\indent As shown in Fig. \ref{fig:OpticalSetup}, our cavity consists of two (M$_{ce}$, M$_{cs}$) identical highly reflective (R = 99.975\%) concave mirrors (Advanced Thin Films) separated by 850~mm. The mirrors have a typical transmittance of 240~ppm (parts per million) and total nominal (scattering and absorption) loss of $<$ 10~ppm. This leads to a maximum power enhancement of $G\approx\sqrt{R}/(1-R)$ =~4,000. The coating is made of alternating dielectric quarter-wave layers of SiO$_2$ and Ta$_2$O$_5$ on a super polished (surface RMS roughness = 0.5~$\AA$) fused silica substrate (thickness 4~mm, diameter 7.75~mm) by an ion beam sputtering technique. The mirrors have a typical radius of curvature of 0.5~m that requires a Gaussian beam waist diameter of 348~$\mu$m at the center of the cavity \cite{SiegmanBook}. This geometry was chosen to maximize the electron-photon collision luminosity when the laser light is frequency locked to the cavity and crosses with the electron beam.\\
\indent A careful study of beam propagation from the frequency doubling setup to the cavity through each optical element has been performed using OptoCad \cite{OCWebsite}. Fig. \ref{fig:BeamTransport} shows the calculated beam size versus the distance along the beam path from the frequency doubling setup. Experimental verification of the laser waist size at the center of the cavity is not possible with the beam profiler. Therefore, we created an auxiliary optical path (2.375 m) that has the same path length from the doubling setup to the cavity with all the optical elements in place. Instead of M$_{ce}$, we used an uncoated mirror substrate made of the same fused silica material with the same radius of curvature. We experimentally checked the beam waist size and location after this mirror substrate, and compared it to our calculation. A small correction with L$_3$ (few mm) is necessary in order to get an average beam size of 350~$\mu$m at a distance of 425~mm (corresponding to the cavity center) from this mirror substrate. At the exit of the cavity, there is a set of optical elements composed of a quarter-wave plate (QWP$_2$), a holographic beam sampler (HBS) and a Wollaston prism (Fig. \ref{fig:OpticalSetup} and Fig. \ref{fig:OpticsTable}) that are used for measuring and monitoring the laser polarization (discussed in Section \ref{sec:LaserPolarization}).
\subsection{Mechanical design of the cavity}
\begin{figure*}[t!]
    \centerline{\includegraphics[width=14.2cm]{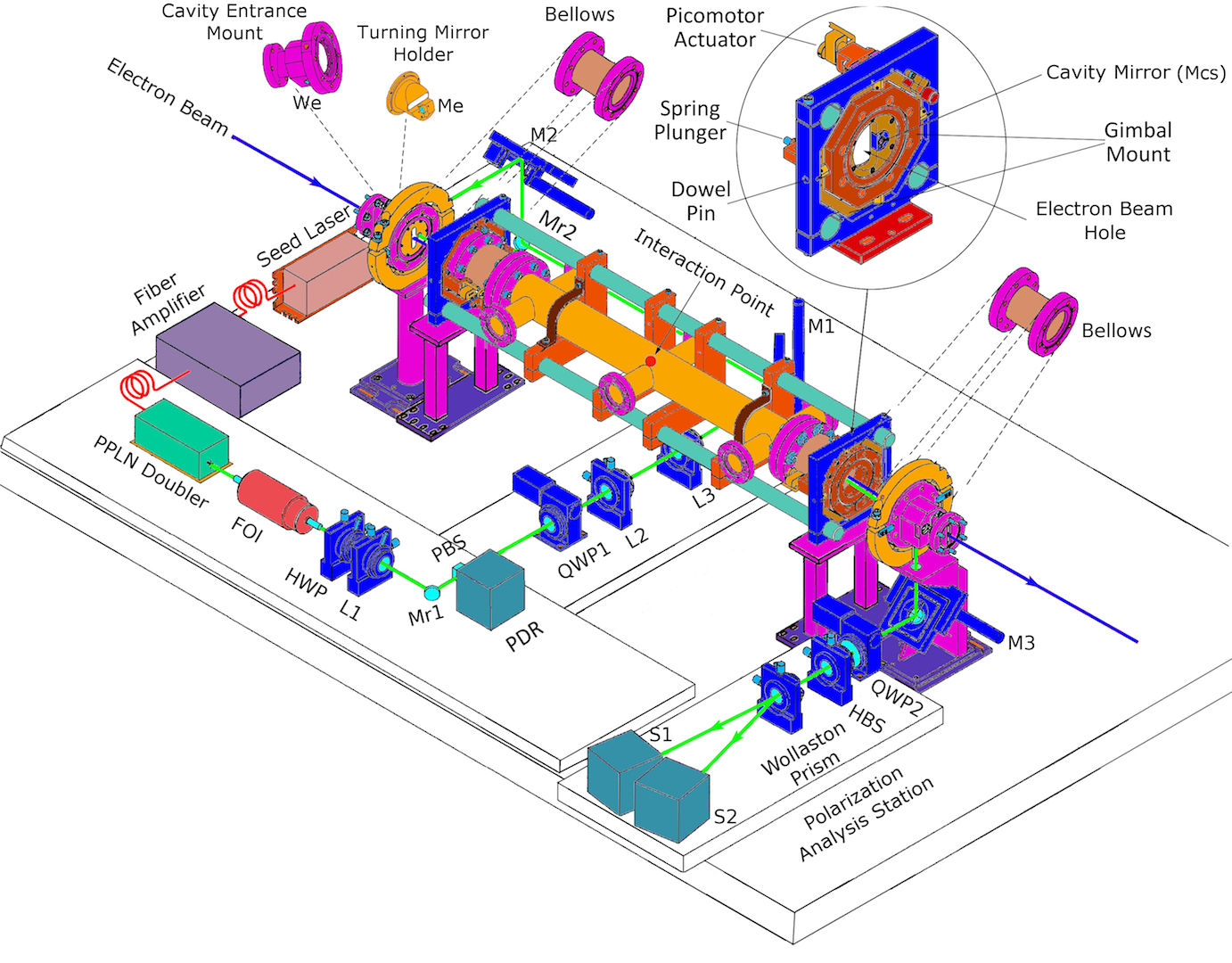}}
    \caption{3D mechanical drawing of the laser optics and Fabry-Perot cavity system in the Hall A tunnel at JLab.}
    \label{fig:OpticsTable}
\end{figure*}
The strongest requirements on the design of a Fabry-Perot cavity in a Compton polarimeter are that it must be rigid and must have a small crossing angle with the electron beam. The small crossing angle increases the electron-photon interaction luminosity. The geometry of the cavity is determined by the total distance between the two mirrors (850~mm), the radius of curvature of the mirrors (500~mm) and the crossing angle between the laser and electron beams. With this design the crossing angle is 24~mrad, the gap between the electron beam and mirror edge is 6.12~mm and the laser beam size on the cavity mirror is 898~$\mu$m.\\
\indent A mechanical drawing of the cavity is shown in Fig. \ref{fig:OpticsTable}. The cavity and all the other optical elements are mounted on an optical table with a footprint of 1.5~m $\times$ 1.2~m. This table is placed on a laminar flow damping system consisting of four pneumatic posts with auto-leveling valves to isolate the vibrations from the ground. The optical table is located inside a small room equipped with a laminar flow fan filter unit. In order to ensure thermal and mechanical stability, an Invar (64FeNi) frame consisting of three cylindrical rods attached to two vertical plates, each with an octagonal cutout, forms the backbone of the cavity frame. The cavity vacuum chamber is a cylindrical vessel made of stainless steel with a diameter of 4.5~inches (Fig. \ref{fig:OpticsTable}) connected to two octagonal gimbal mounts through two soft bellows. The bellows allow the gimbal mounts to tilt freely in both the horizontal and vertical planes when they are adjusted by four piezo actuators under atmosphere. However, in our experiment, the bellows were too stiff under vacuum for the piezo actuators to move the gimbals remotely. Therefore the procedure involves locking the cavity under atmosphere, then locking the gimbals and drawing vacuum\footnote{ Even though we were able to keep the cavity alignment in position during the three-month period of the PREx experiment, we believe this feature could be redesigned to make the remote alignment of cavity mirrors under vacuum possible.}. The cavity mirror is mounted on a mirror holder attached to these gimbal mounts which are also made of Invar. Each gimbal mount is supported by four stainless steel cylindrical bearings (C-Flex Bearing Co.) that form two axes (horizontal and vertical) for each gimbal mount and allow them to tilt freely around each axis. The mirror holder is machined such that the rotation axis of each mirror lies on the same axis as the gimbals. The bearings are 0.25~inch in diameter and 0.4~inch in length, and each can support a load up to 100~lbs. Two remote controlled picomotors (8302; Newfocus) attached to each gimbal mount are used for aligning each cavity mirrors by tilting the gimbal mounts in both the vertical and transverse planes with respect to the laser beam propagation direction. A pair of counteracting spring plungers (maximum load 13~lbs) are attached to the gimbal mounts to keep the alignment in position. The picomotors are interfaced to the EPICS \cite{EPICS} slow control system (discussed in Section \ref{sec:feedback}) that allows remote alignment of the cavity mirrors.\\
\begin{figure}[t!]
    \centerline{\includegraphics[width=7.2cm]{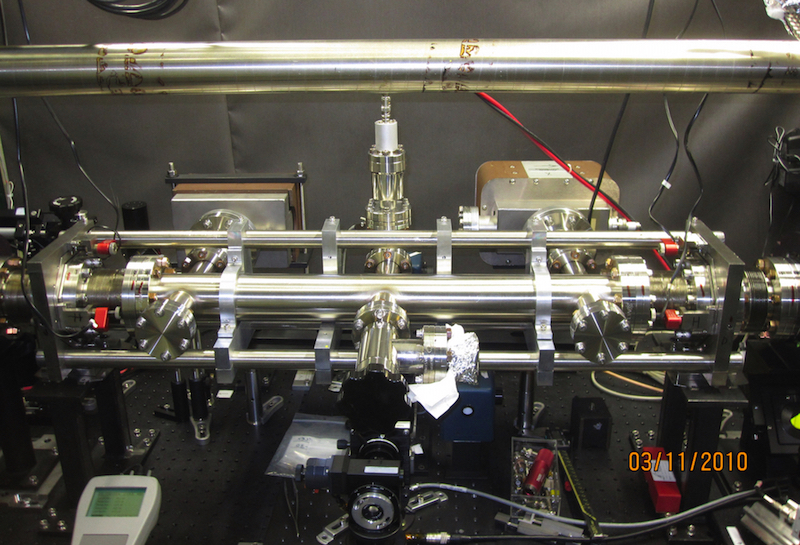}}
    \caption{The cavity vacuum chamber attached to two ion pumps on the optical table and connected to the electron beam pipe in the Hall A tunnel at JLab. The electron beam pipe above the cavity is used for a straight beam to the experimental Hall when the Compton polarimeter is not used.}
    \label{fig:CavityInstalledHall}
\end{figure}
\indent A pair of vacuum window substrates (W$_e$ and W$_s$ in Fig. \ref{fig:OpticalSetup}) that are made of fused silica (3~mm thick and 0.7 inch in diameter) allow the laser beam to enter and exit the cavity via a pair of 0.5~inch turning mirrors (M$_{e}$ and M$_{s}$ in Fig. \ref{fig:OpticalSetup}) oriented at 45$^{\circ}$ with respect to the incident laser beam. The vacuum windows are AR coated for 532~nm and welded to stainless steel flanges (Fig. \ref{fig:OpticsTable}) by the glass-metal soldering technique. Two soft bellows on both sides of each gimbal mount are connected to both sides of the cavity vacuum chamber through a pair of flanges and allow the gimbals to tilt freely. Each of the 45$^{\circ}$ turning mirrors is mounted to an aluminum holder that is attached to a stainless steel flange mounted on a post. Each aluminum holder has a slit 4 cm long and 1cm wide that allows the electron beam to pass through. When the cavity is installed in the electron beam line (Fig. \ref{fig:CavityInstalledHall}), the stainless steel flanges are connected to the beam pipe by another set of soft bellows that isolates any vibrations from the rest of the electron beam pipe. Two beam position monitors located on both sides of the cavity are used to monitor changes in beam position of the electron beam during beam tuning and data taking. Two ion pumps can provide a vacuum of $\sim$ 10$^{-9}$~Torr inside the cavity after several hours of bake-out at $\sim$ 100~$^\circ$C.
\subsection{Feedback and slow control system}\label{sec:feedback}
\begin{figure*}[t!]
    \centerline{\includegraphics[width=12.5cm]{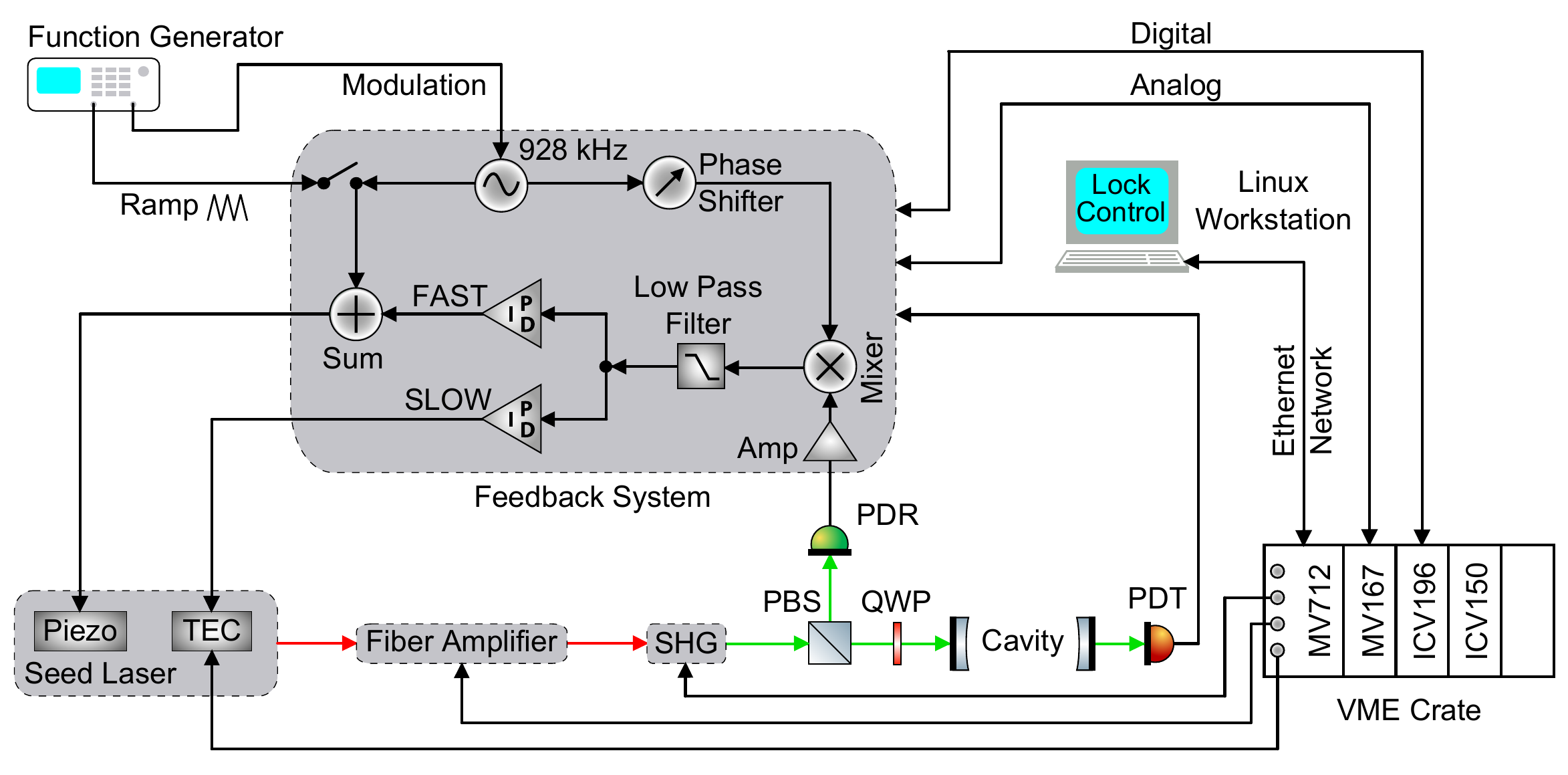}}
    \caption{A simplified flow chart of the feedback control system used for laser frequency locking to the cavity fundamental mode.}
    \label{fig:CavContrl}
\end{figure*}
\begin{figure*}[b!]
    \centerline{\includegraphics[width=15.7cm]{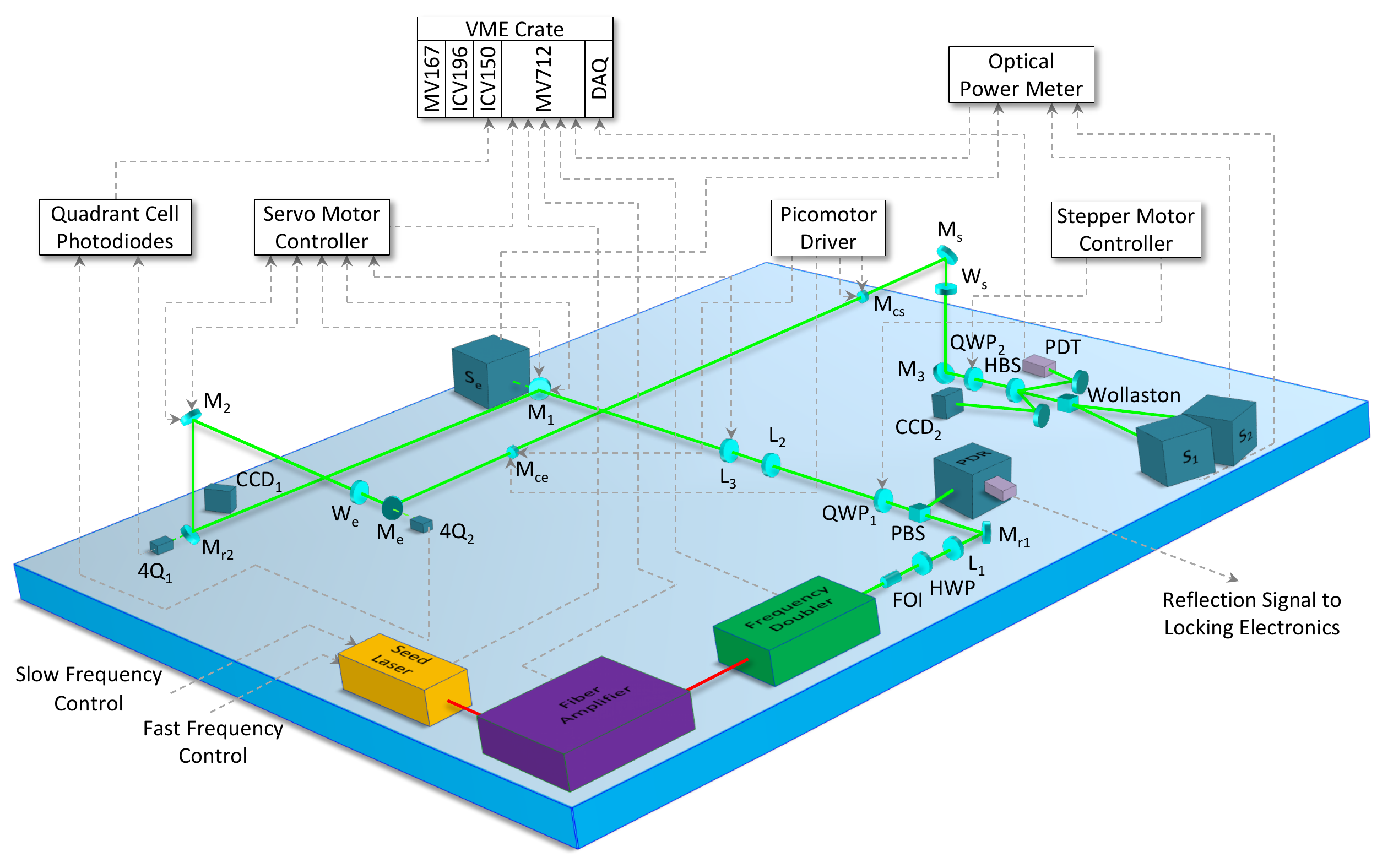}}
    \caption{A functional view of the laser and cavity system illustrates the electronic control units with the optical elements.}
    \label{fig:CavityFunctionalView}
\end{figure*}

Maintaining the resonance and therefore the enhancement in the cavity requires feedback control of the laser frequency. The optical frequency stabilization must be better than the free spectral range ($FSR$=176.5~MHz) divided by the finesse ($F$=12,000), which is $\sim$15~kHz (cavity bandwidth). This is equivalent to a stabilization of the 850~mm cavity length to better than the laser wavelength divided by the finesse, which is $\sim$ 0.04~nm. The electronic feedback system that controls the laser frequency uses the Pound-Drever-Hall technique \cite{PDH}. The error signal is created by phase detection between the photodiode (PDR in Fig. \ref{fig:CavContrl}) output of the modulated reflection signal and the modulation signal from the function generator (FG220, Yokogawa). The error signal includes information on cavity resonance, and is then used to build the feedback signals supplied on both the fast and slow control ports of the seed laser. The electronic feedback system has been designed and built by Saclay \cite{JPJorda}. The simplified flow chart is shown in Fig. \ref{fig:CavContrl}.\\
\indent The optical elements that are critical to beam alignment and polarization control and measurement, the lasers (seed laser, fiber amplifier and PPLN doubling system) and the locking electronics are all interfaced to a remotely controlled system through a local workstation. Electronics specific to the feedback control loop, the function generator (ramp generator, modulation signal generator), the oscilloscope and the workstation are located in the control room. The lasers, stepper and servo motors with their control units, photodiodes, CCD cameras and the preamplifiers are located on the optical table in the Hall A tunnel. The monitors related to the cavity resonance mode and beam alignment are connected to cameras CCD$_2$ and CCD$_1$ through coaxial BNC cables and are located in the control room. Fig. \ref{fig:CavityFunctionalView} shows a functional view of the laser and cavity system with control units.\\
\indent An automatic switching system from ``open-loop" mode to ``closed-loop" mode around the cavity resonance region permits the system to transition automatically between the resonant and locked states. It consists of an electronic circuit and an EPICS program that manage the laser temperature scan via the slow control port of the seed laser. A 10~V peak-to-peak triangular ramp, together with a sinusoidal modulation signal at $\Omega$ = 928~kHz with 50~mV amplitude, is supplied on the laser piezoelectric transducer (PZT) via the fast control port of the seed laser.\\
\indent Two fast photodiodes (S1223; Hamamatsu) are used for monitoring the reflected (PDR) and transmitted (PDT) signals. They are held at a constant voltage level of 5~V. The currents from the photodiodes are transformed into voltage signals via a trans-impedance amplifier that allows signal transmission across 100 meters of coaxial cables. A band-pass filter is applied on the reflected signal from the cavity after it is mixed with the modulation signal at frequency $\Omega$ and amplified. The mixing and filtering extracts an error signal corresponding to the change in reflected light at the laser modulation frequency. The value of $\Omega$ was determined by minimizing the laser Residual Amplitude Modulation (RAM) \cite{JPJorda}. The error signal is injected into a series of three separate integrators (proportional-integral-derivative or PID) common to the slow and fast control loops. The two control modes play complementary roles: the slow mode is for compensating the slow drift in laser frequency while the fast mode allows the efficient reduction of the laser frequency jitter. The output signals of these two modules are applied directly to the two laser control ports that control the seed laser frequency.\\
\indent The VME (VERSA Module Eurocard) crate, used for controlling the electronics, is located in experimental Hall A. The electronics are controlled from a workstation in the control room with the help of interface cards. The interface cards permit us to transmit numerical signals over a network to the optical table area which is about 100 meters from where the VME crate is located. The EPICS slow control program is used for regulating and activating the entire system from the workstation in the control room. It allows real-time control of analog and digital inputs and outputs to the various crates and modules. The VME crate houses the following cards: an ICV150 card that measures the voltages of the control signals (PDR, PDT, fast ramp, slow ramp); an ICV196 card that provides the digital interface between the electronic-card sequencer and the control screens by sending and receiving transistor-transistor logic (TTL) signals; and an MV712 card that allows RS-232 commands to control devices such as the laser, the PPLN temperature controller and the fiber amplifier. A servo motor controller is used for controlling the steering mirrors M$_1$, M$_2$ and the focusing lens L$_3$. Beam position variations are monitored with two quadrant cell photodiodes, noted as 4Q$_{1}$ and 4Q$_{2}$, that detect a small amount of transmitted light ($< 0.1\%$) behind M$_{r2}$ and M$_e$. The analog signals read by their electronics are fed to the ICV150 card and monitored by the control system. Each quarter wave plate (QWP$_1$, QWP$_2$) is controlled by a separate stepper motor controller (KS491-2P, Suruga Seiki). With the help of the slow control interface, we can control the entire locking assembly from the workstation: the locking threshold level, the servo loop gain, the state of the feedback loop and the laser frequency scan program.
\subsection{Cavity performance}
\begin{figure}[t!]
    \centerline{\includegraphics[width=7.8cm]{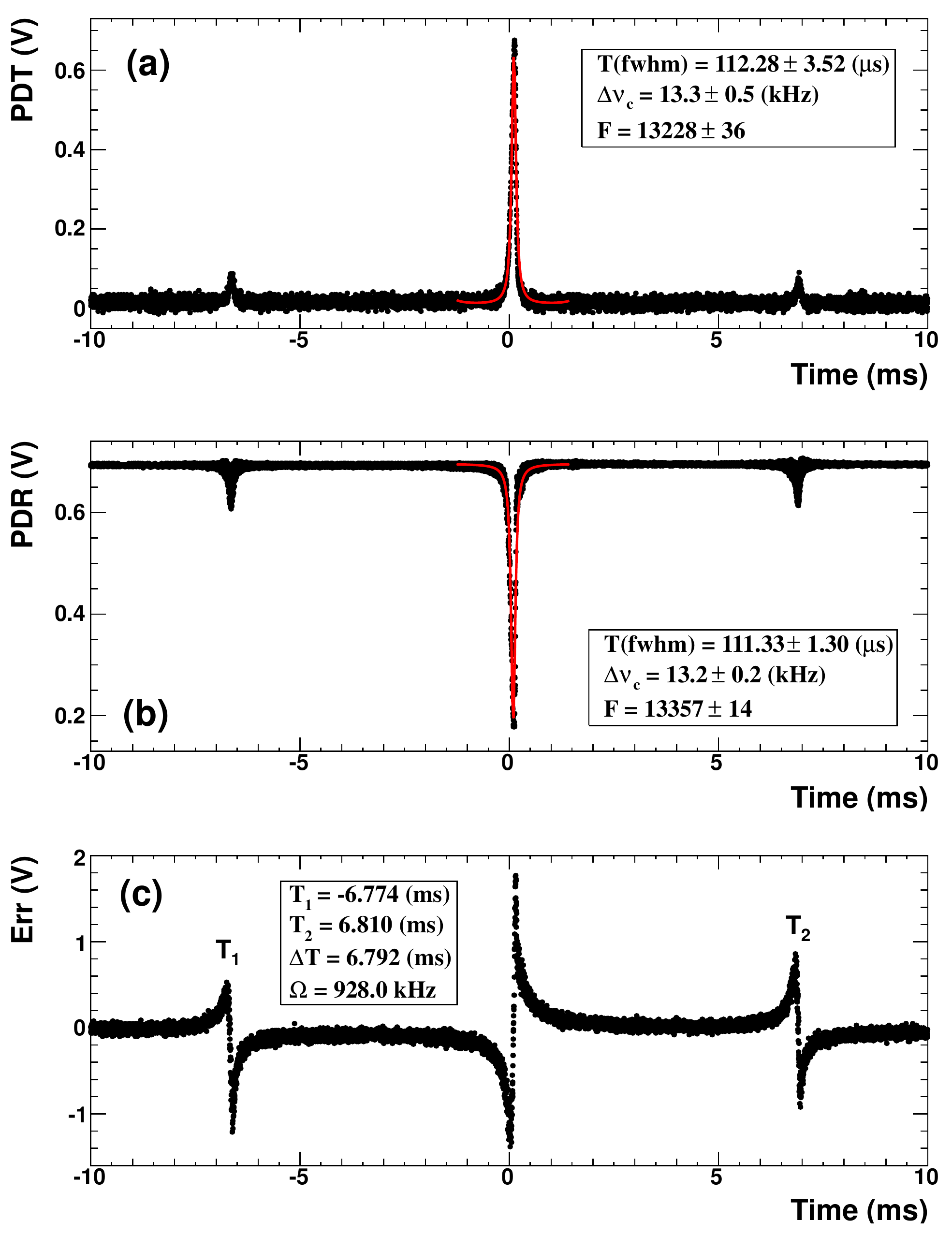}}
    \caption{A theoretical fit \cite{SiegmanBook} (red line) to the transmission (a) and reflection (b) signals used to extract the cavity bandwidth when the cavity is on resonance. A typical error signal (c) of the PDH feedback control is also used to convert the time units to the frequency units in order to calculate the cavity finesse.}
    \label{fig:Bandwidth}
\end{figure}
Once the optical cavity is installed in the electron beam line, it is important to check various parameters (such as cavity finesse $F$, bandwidth $\Delta \nu_{c}$ and the intra-cavity power $P_{c}$) that are essential to the performance of an operational cavity. Two methods have been employed to characterize the cavity: a graphical method and a decay-time measurement method.\\
\indent The graphical method analyzes the shape of the cavity resonance waveforms (transmission, reflection and error signals) while the seed laser frequency is being scanned by a triangular ramp voltage applied to the laser PZT. In this method, we first determine the FWHM of the reflection/transmission peak that corresponds to the cavity bandwidth. Then, one converts from time units to frequency units by setting the gap, 2$\Delta T$ = $|T_1|$ + $|T_2|$, between the two sidebands ($T_1$, $T_2$) equal to twice the modulation frequency $\Omega$. Fig. \ref{fig:Bandwidth} shows a theoretical fit \cite{SiegmanBook} to the transmission (Fig. \ref{fig:Bandwidth}(a)) and reflection (Fig. \ref{fig:Bandwidth}(b)) signals that are used to extract the cavity bandwidth and therefore the cavity finesse. The measurement shows that the cavity exhibits a typical bandwidth of $\sim$ 13~kHz with a corresponding finesse of $\sim$ 13,300. Since the typical width of the resonance is $\sim$ 0.04~nm, the measurement is susceptible to external disturbances and fluctuates for each measurement. In addition, due to a hysteresis exhibited in the laser PZT, the modulation sidebands would not be symmetric around the main peak as shown in Fig. \ref{fig:Bandwidth}. Therefore, the finesse measured in this method has a large error at the order of 10$\%$ -- 20$\%$.\\
\begin{figure}[t!]
    \centerline{\includegraphics[width=7.8cm]{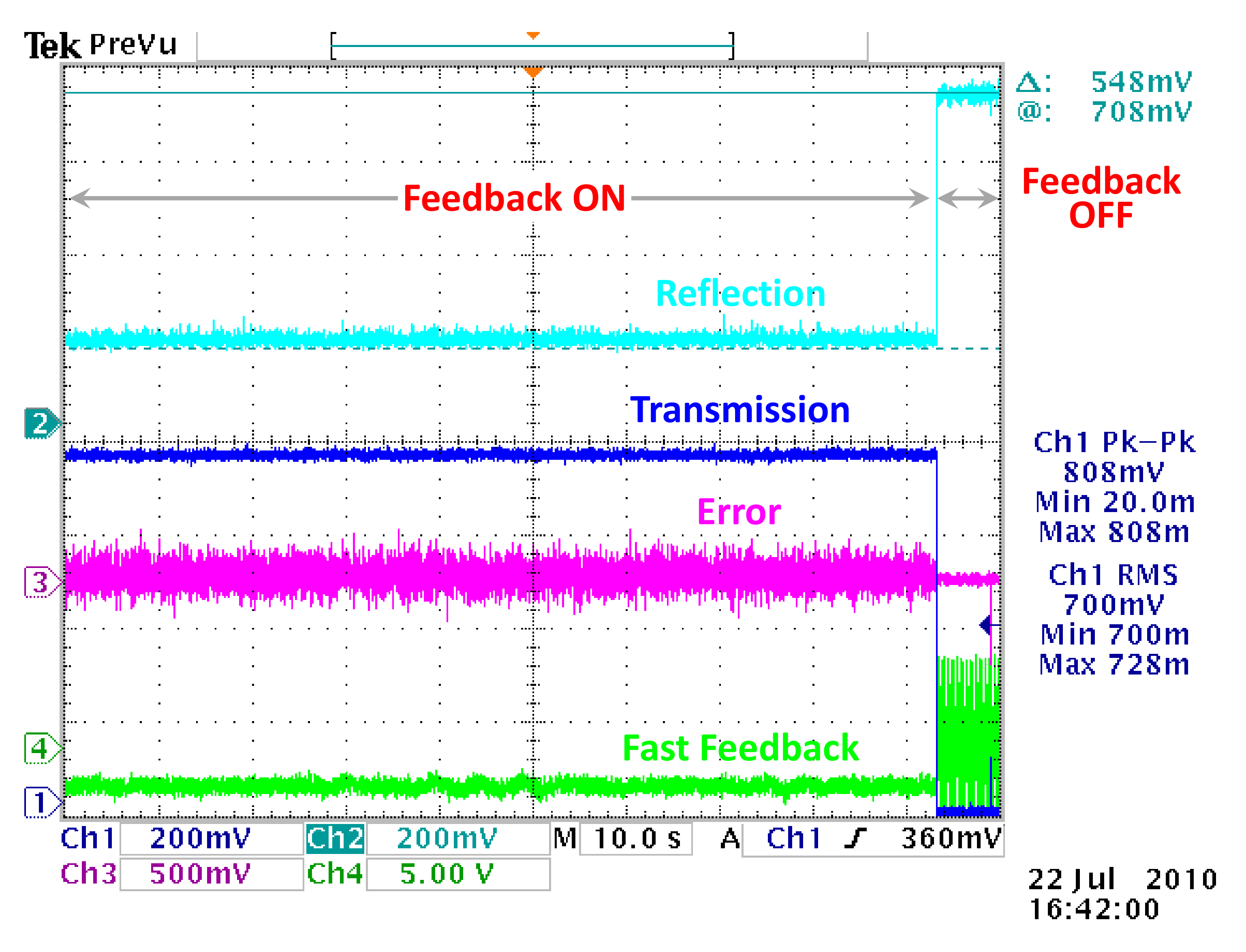}}
    \caption{The digital oscilloscope traces are, from top to bottom, power reflected at the entrance of the cavity (cyan), the power transmitted by the cavity (blue), the PDH error signal (magenta) used for generating the fast feedback signal (green) that is given to the laser PZT. The trace shows cavity locking signals corresponding to locked and unlocked states of the cavity.}
    \label{fig:ScopeSignal}
\end{figure}
\indent Once the laser beam and cavity are aligned with respect to each other, the laser frequency can be locked to the fundamental cavity resonance mode via its feedback system. Fig. \ref{fig:ScopeSignal} shows a digital oscilloscope trace that includes waveforms corresponding to the locked (feedback on) and unlocked (feedback off) states of the cavity. In the cavity decay-time measurement method, while the cavity is locked, we switch off the laser diode and immediately record the transmitted power via a fast photodiode (S1223; Hamamatsu) and a digital oscilloscope (TDS3054B; Tektronix). For a high-finesse cavity, the cavity decay time $T_d$ is directly related to the cavity parameters by \cite{VeryHighQCavityRSI, RaminOL},
\begin{equation}\label{eq:CavityDecay}
T_d \approx \frac{F L_{c}}{\pi c} = \frac{1}{2 \pi \Delta \nu_c},
\end{equation}
where $L_c$ is the cavity length and $c$ is the speed of light. Here, it is necessary to take into account the decay time of the laser diode ($T_{l}$) itself in order to determine $T_{d}$ correctly.
\begin{figure}[t!]
    \centerline{\includegraphics[width=7.5cm]{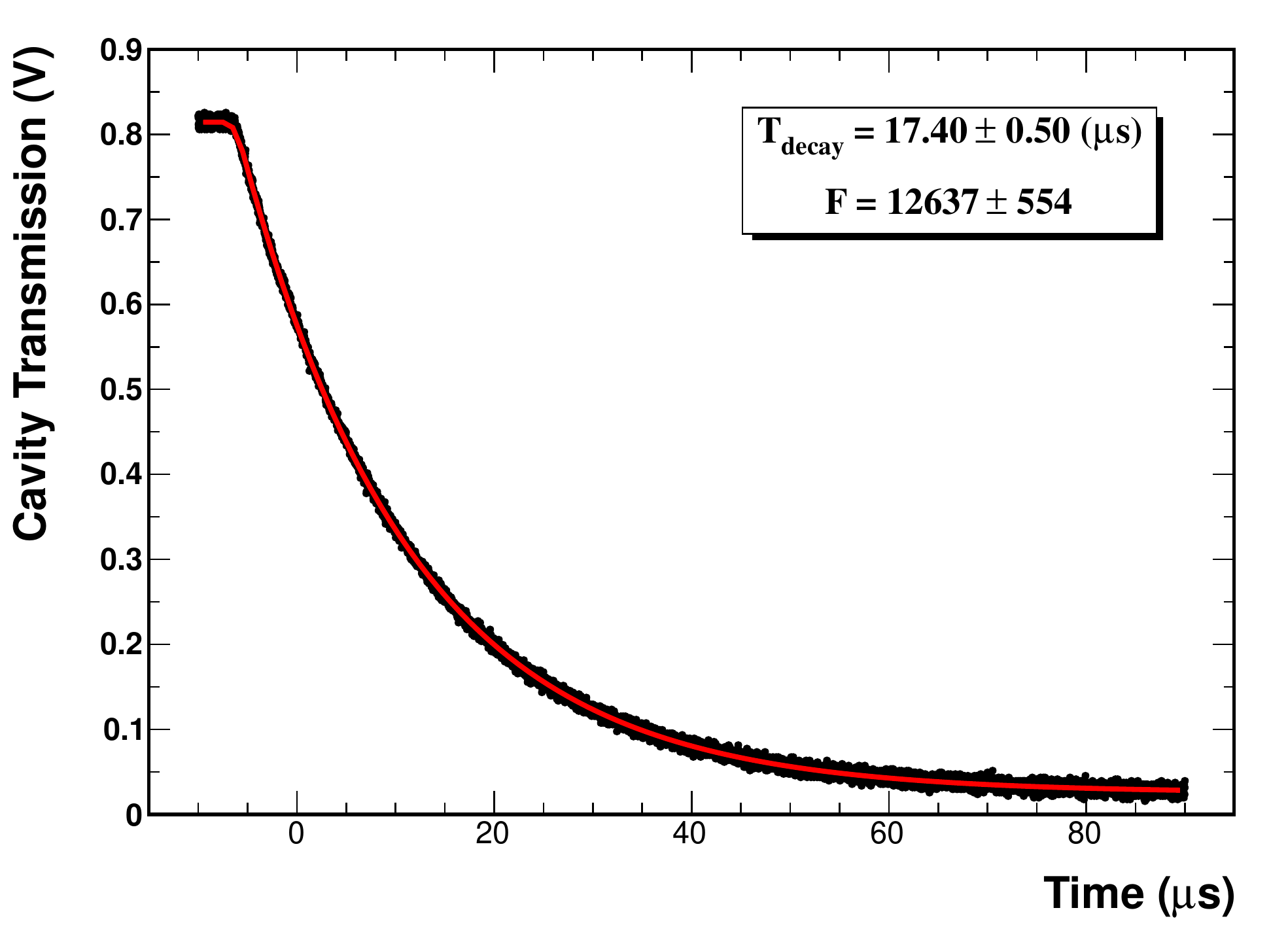}}
    \caption{Decay curve of the intra-cavity power. The exponential curve \cite{RaminOL} (red line) is fitted to the transmitted power of the cavity (black dots) to extract the cavity decay time. The finesse is corrected for the laser diode decay time of 6~$\mu$s.}
    \label{fig:DecayTime}
\end{figure}
Fig. \ref{fig:DecayTime} shows a typical cavity decay curve measured (black dots) for the optical cavity and an exponential fit \cite{RaminOL} (red line) to extract the decay time. Here, after correcting the fitted value of the decay time to the measured laser decay time (6~$\mu$s) \cite{NFalletto} and taking the sampling error of the oscilloscope as 0.5~$\mu$s, we have $T_d$ = 11.40 $\pm$~0.5~$\mu$s with the corresponding $F$ = 12,637 $\pm$~554 and $\Delta\nu_c$ = 14.00 $\pm$~0.32~kHz from this measurement.\\
\begin{table*}[b!]
\caption{Characterization of the cavity parameters for a typical run during the PREx experiment.}\label{Table: Characterization of the cavity parameters during PREx.}
\begin{center}
\begin{tabular}{  l  l  l  l  }    \hline
~Parameters ~~~~~~~~~~~& Symbol~~~~~~~& Value~~~~~~~~~& Unit  \\\hline
~Incident Laser Power: &~$P_i$~~~~~~~&    1.24 $\pm$ 0.012~~~~~~~~~& W\\
~Transmitted Power: &~$P_t$	~~~~~~~&    0.75 $\pm$ 0.007~~~~~~~~~& W\\
~Cavity Decay Time: &~$T_{d}$~~~~~~~&    11.40 $\pm$ 0.50~~~~~~~~~& $\mu$s\\
~Finesse: &~$F$	~~~~~~~&    12600 $\pm$ 550~~~~~~~~~& \\
~Cavity Bandwidth: &~$\Delta \nu_c$~~~~~~~&  14.00 $\pm$ 0.32~~~~~~~~~& kHz\\
~Optical Coupling: &~$\alpha$	~~~~~~~&   0.79 $\pm$ 0.08~~~~~~~~~& \\
~Loss Ratio: &~$\beta$	~~~~~~~&    0.14 $\pm$ 0.013~~~~~~~~~& \\
~Mirror Transmittance: &~$T$ = $\dfrac{\pi}{F (1 + \beta)}$~~~~~~~&   200 $\pm$ 9~~~~~~~~~& ppm\\
~Mirror Losses: &~$\gamma$ = $\dfrac{\pi \beta}{F (1 + \beta)}$	~~~~~~~&   30 $\pm$ 3~~~~~~~~~& ppm\\
~Maximum Enhancement: &~$G_{\text{max}}$ = $\dfrac{T}{(\gamma + T)^2}$	~~~~~~~&   3800 $\pm$ 170~~~~~~~~~& \\
~Intra-cavity Power: &~$P_c$ = $\alpha G_{\text{max}} P_i$~~~~~~~&   3750 $\pm$ 120~~~~~~~~~& W\\
\hline            
\end{tabular}
\end{center}
\end{table*}
\indent Now that $T_{d}$ and $\Delta \nu_c$ are determined, the maximum cavity enhancement ($G_{\text{max}}$), optical coupling of laser fundamental mode to the cavity ($\alpha$) and mirror characteristics such as loss ($\gamma$) and transmittance ($T$) need to be calculated to determine the intra-cavity power. Based on the theoretical models described in Refs. \cite{NFalletto, CHoodPRA, ShernoffAO}, cavity parameters can be determined from several external measurements such as the power going into the cavity ($P_{i}$), the power transmitted out of the cavity ($P_{t}$) and the power reflected from the cavity ($P_{r}$). If we ignore the power contained in modulation sidebands, the loss ratio ($\beta = \gamma/T$) of the mirrors can be written as,
\begin{equation}
\beta = \frac{1}{2} \Bigg[ \frac{P_{i}}{P_{t}} \Bigg(1- \frac{P_r^{\mathit{\text{on}}}}{P_r^{\mathit{\text{off}}}} \Bigg) - 1 \Bigg],
\end{equation}
where $P_r^{\text{on}}$ is the reflected power measured when the cavity is locked, and $P_r^{\text{off}}$ is the reflected power measured when the cavity is unlocked. Now the optical coupling $\alpha$ is related to $\beta$ by,
\vspace{0.2cm}
\begin{equation}
\alpha = \frac{P_{t}}{P_{i}} (1 + \beta)^2
\end{equation} 
According to Fig. \ref{fig:ScopeSignal}, and measured values of $P_{i}$ and $P_{t}$, we can calculate $\beta$ and $\alpha$, which will allow us to calculate $T$, $\gamma$, $G_{\text{max}}$ and $P_{c}$. Table \ref{Table: Characterization of the cavity parameters during PREx.} summarizes the cavity parameters measured during the PREx experiment. The stored power in the cavity is estimated to be $\sim$ 3.7~kW with a corresponding enhancement of $\sim$ 3,800. As seen in Fig. \ref{fig:ScopeSignal}, the reflected and transmitted power from the cavity is extremely stable after turning on the feedback system, suggesting that there is no thermal effect due to high power photons incident on the cavity mirrors during an extended period of time.\\
\indent We have described the optical, mechanical and electronic systems of our setup and have discussed the results obtained during the PREx experiment. Robust cavity locking with stable high power is essential to achieve a good signal-to-noise ratio, but in order to observe Compton scattering asymmetry and therefore measure the electron beam polarization, we have to create a highly circularly polarized photon beam at the CIP. In the following section, we describe the polarization aspects of our system.
\section{Intra-cavity laser polarization}\label{sec:LaserPolarization}
In terms of polarization, our optical setup needs to fulfill several functions such as creation of a highly circularly polarized photon beam at the CIP, switching of left- and right-circular polarization at regular intervals and monitoring of the laser beam polarization at the CIP in situ by using a laser polarization transfer function.
\subsection{Circularly polarized light}\label{sec:CircPol}
As we described in Section \ref{sec:FPCavity}, the IR beam from YDFA is vertically polarized (extinction ratio: 20~dB), as is the green beam after the PPLN doubler ($99.88\%$). As shown in Fig. \ref{fig:OpticalSetup}, a well collimated beam ensures the purity of polarization states after it passes through various polarization-shaping elements, such as HWP, PBS and QWP$_1$. After passing through the FOI, HWP and PBS (extinction ratio: 27~dB), the polarization of the green beam is horizontal before it is converted to circular (left/right) by QWP$_1$ (zero-order), which is mounted on a stepper motor. The stepper motor reverses the helicity of the circular polarization at a time interval of 40 seconds.\\
\indent The system must provide the highest degree of circular polarization at the CIP. It is crucial not only for a high experimental asymmetry but also to achieve a good precision in the electron beam polarization measurement. In our setup, the polarization state is controlled by the PBS and QWP$_1$. We have a circularly polarized beam generated after QWP$_1$ and the signal reflected by the cavity can be separated from the incident beam after the PBS and detected by a fast photodiode (PDR). However, we must take into account the unavoidable degradation of the polarization (birefringence) between the output of the QWP$_1$ and the CIP. In our setup, the main source of birefringence is the dielectric steering mirrors after QWP$_1$ which have a difference of $\sim$ 0.5\% between the reflection coefficients for $S$ and $P$ polarizations at 45$^{\circ}$. This effect can be reduced by using a well established compensated mirror scheme for laser polarization transport \cite{SLDCompton, NFalletto}. In this scheme, two pairs of identical 45$^{\circ}$ dielectric mirrors (M$_{1}$ -- M$_{r2}$ and M$_{2}$ -- M$_{e}$) are oriented at the same angle of incidence, but with perpendicular incident planes. In this way, the $S$-wave at the first mirror becomes the $P$-wave at the second mirror. If the mirrors are identical, then the difference in reflectivity and phase between the components is canceled after the last mirror (M$_{e}$). With this scheme, without the cavity mirrors in place, we obtained a maximum left circular polarization of 99.57\% for a QWP$_1$ angle of -50$^{\circ}$ (counter-clockwise) and a maximum right circular polarization of -98.07\% for a QWP$_1$ angle of 50$^{\circ}$ (clockwise) at the CIP. In this optimization, the initial polarization state was also adjusted by tilting the QWP$_1$ slightly off normal to the beam. Table \ref{Table: Measurement of $DOCP$} summarizes the degree of circular polarization ($DOCP$) measurement after QWP$_1$ and at the CIP without cavity mirrors in place. Here, we believe the measured asymmetry of 1.5\% between the two polarization states at the CIP is due to manufacturing variations in the compensating mirror coatings.\\
\begin{table}[b!]
\caption{Measurement of the degree of circular polarization ($DOCP$) after QWP$_1$ and at the CIP without cavity mirrors.}\label{Table: Measurement of $DOCP$}
\begin{center}
\begin{tabular}{ c | c}    \hline
Location 				& 	{$DOCP$ (\%)}\\\hline
exit of QWP$_1$ 		& 	99.96 $\pm$ 0.1 (Left)\\
  					&  	-99.98 $\pm$ 0.1 (Right)\\\hline
at the CIP 			& 	99.57 $\pm$ 0.1 (Left)\\
(without cavity mirrors) 	&  	-98.07 $\pm$ 0.1 (Right)\\\hline   
\end{tabular}
\end{center}
\end{table}
\indent The highly circularly polarized laser beam at the CIP, must be monitored during the data taking of the physics experiment. Since intra-cavity parameters are inaccessible from outside, this involves mapping a precise polarization transfer function that can extract the intra-cavity polarization from the external polarization measurement.
\subsection{Laser polarization transfer function}
\begin{figure}[t!]
    \centerline{\includegraphics[width=7.8cm]{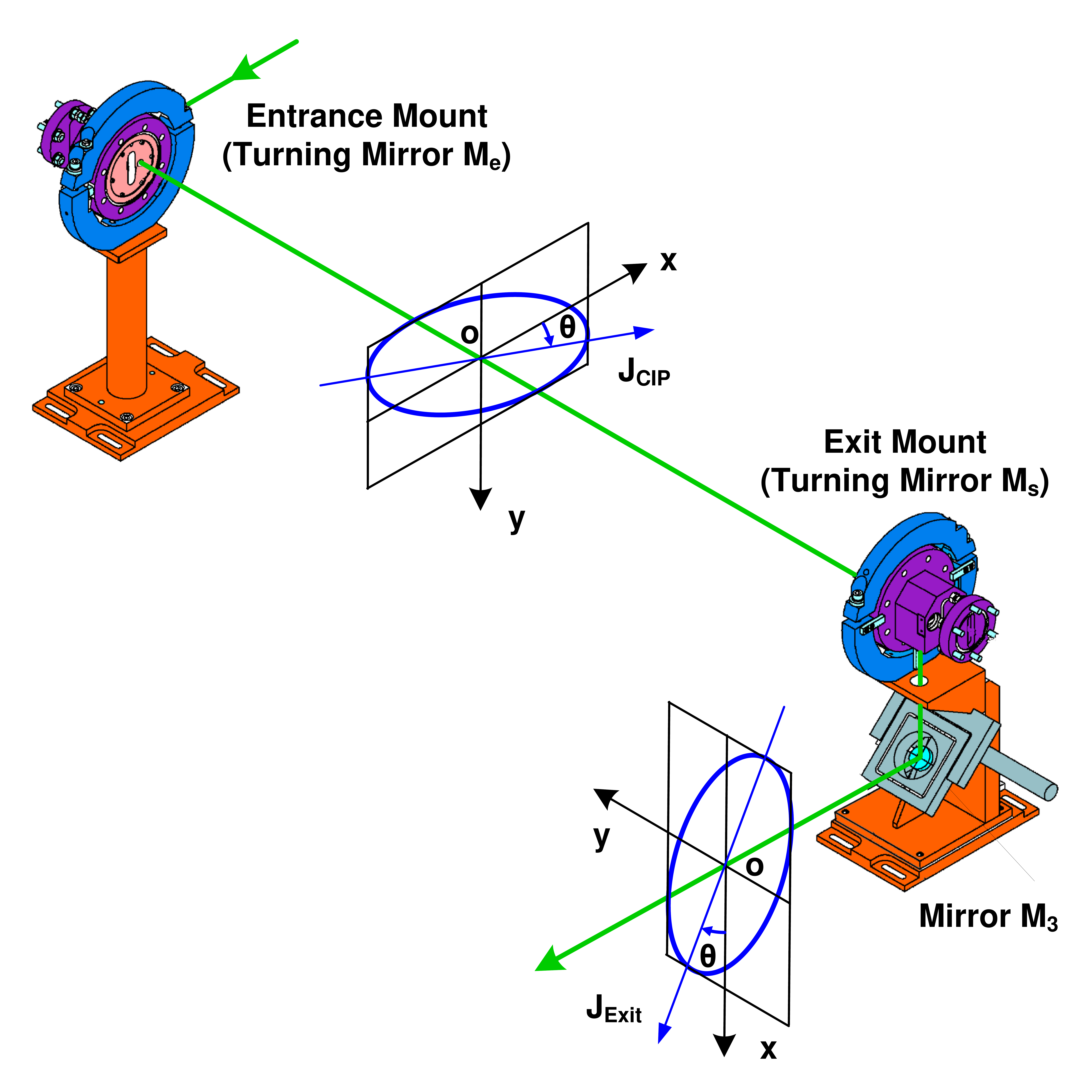}}
    \caption{Propagation of polarization ellipse from the CIP to the cavity exit line. The schematic illustrates a case in which the cavity between the two stands is removed.}
    \label{fig:TF_Measure}
\end{figure}
The principle of our approach is to characterize the polarization state at the center of the cavity for a measured state of polarization at the cavity exit. The transfer function gives full information about the elements of the Jones matrix \cite{Jones1, Jones2} for a given optical system based on well known initial and final states, thus one can propagate a polarized beam through a system and calculate its final state. Similarly, a final polarization state can be back-propagated to its initial state. In our case, the transfer function allows us to determine the polarization and its orientation at the cavity center (i.e. CIP) based on the degree and orientation of the polarization measured at the cavity exit. The system we have to model is composed of two mirrors M$_{s}$ and M$_{3}$ (Fig. \ref{fig:TF_Measure}). The Jones vector representing the polarization state at the cavity center $\hat{J}_{\text{CIP}}$ is related to the exit polarization $\hat{J}_{\text{Exit}}$ using,
\begin{equation}
\hat{J}_{\text{Exit}} = [TF]^{-1}  ~ \hat{J}_{\text{CIP}},
\end{equation}
where $[TF]$ represents the transfer function between them. $[TF]$ is a 2 $\times$ 2 complex matrix which includes information about the phase shift $\delta$ upon reflection on the mirror, the polarization orientation rotation $\theta$ introduced by a mirror with respect to the axis $Ox$ (Fig. \ref{fig:TF_Measure}) and another rotation angle $\phi$ caused by any change of orientation of the coordinate frame. The characteristic transfer matrix of a dielectric mirror is represented by the product of associated matrices \cite{SalehTeichBook},
\begin{equation}
M (\delta, \theta, \phi) = P(-\theta)~R (\delta)~P(\theta)~T (\phi),
\end{equation}
and the total transfer function of a system composed of two mirrors (M$_{s}$ and M$_{3}$ in Fig. \ref{fig:TF_Measure}) is,
\begin{equation}\label{eq:TF}
[TF] = [M_s(\delta_s, \theta_s, \phi_s) ~ M_3(\delta_3, \theta_3, \phi_3)],
\end{equation}
and $\hat{J}_{\text{CIP}}$ is calculated by,
\begin{equation}
\hat{J}_{\text{CIP}} = [TF] ~ \hat{J}_{\text{Exit}}
\end{equation}
As seen from Eq. (\ref{eq:TF}), for a system with two dielectric mirrors, six parameters characterize its full transfer function. We begin by preparing a set of polarization states with a well known $DOCP$ and ellipse angle at the CIP and then measure the corresponding polarization state at the cavity exit. Once we have the initial and final state vectors based on these states with orientations covering the range from 0$^{\circ}$ to 180$^{\circ}$, theoretically we have enough equations to solve and determine the parameters.\\
\begin{figure*}[t!]
\centering
    \centerline{\includegraphics[width=14.5cm]{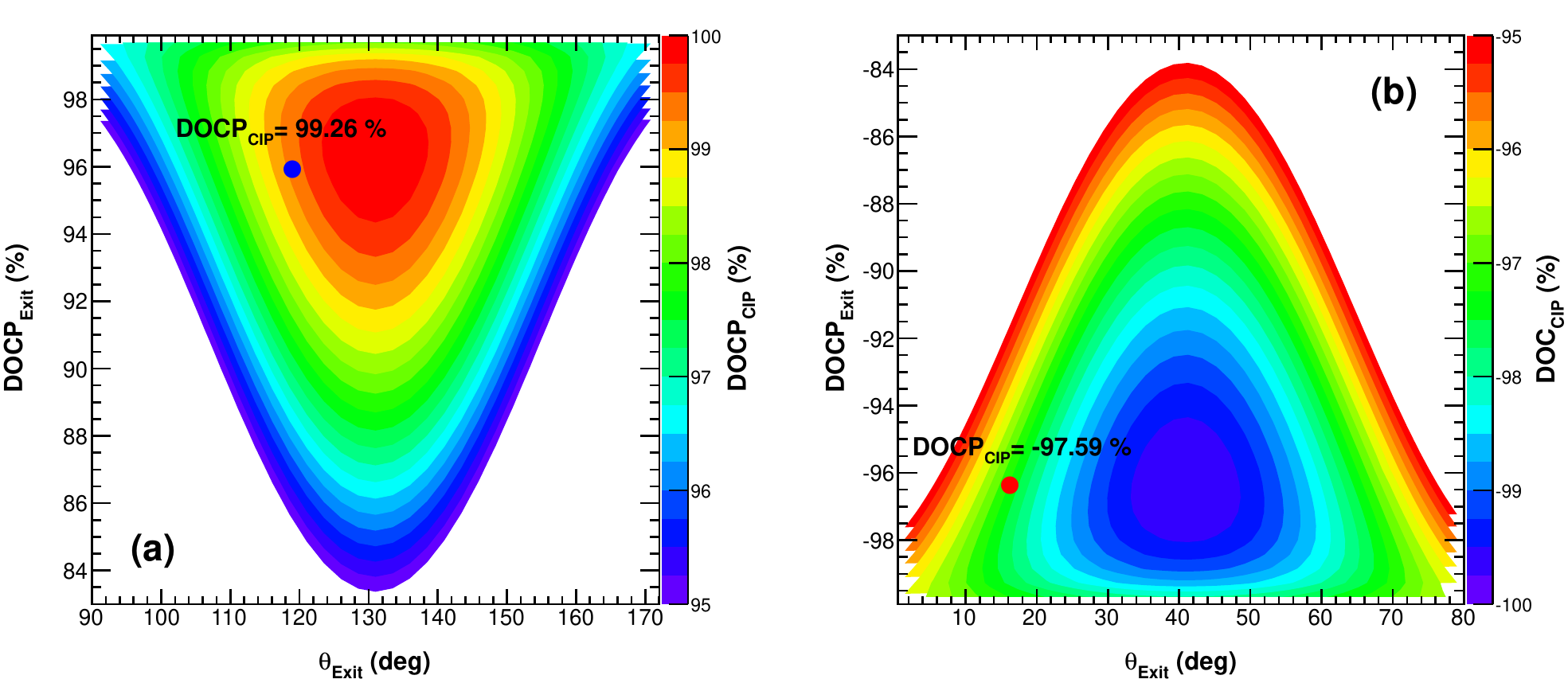}}
    \caption{A contour map of the transfer function for the left (a) and right (b) circularly polarized states of the CIP with respect to the exit polarization states. The Exit $DOCP$ was zoomed in to between $\pm$83\% to $\pm$100\% for illustration purposes. The numbers corresponding to blue (a) and red (b) dots in figures are the values of nominal $DOCP$ at the CIP based on exit polarization measurements that are also shown in Table \ref{Table: CIPdocp}.}
    \label{fig:TFCounterview}
\end{figure*}
\indent The experimental determination of the polarization transfer function involves breaking the cavity vacuum, removing the cavity mirrors and conducting a set of polarization measurements both at the CIP and cavity exit as shown in Fig. \ref{fig:TF_Measure}. A station that is composed of a Glan-Laser polarizer (extinction ratio: 50~dB) and a quarter-wave plate (zero-order) is used for generating a set of polarization eigenstates at the CIP. When we have a given polarization state with a corresponding $\theta$, we rotate both elements (Glan-Laser polarizer and quarter-wave plate) by the same angle which leads to a rotation of $\theta$ for a fixed degree of polarization. We have started with light of $DOCP$ = $\pm$92\% with various orientations (10$^{\circ}$ -- 170$^{\circ}$) at the CIP by rotating both the polarizer and the quarter-wave plate with a step size of 20$^{\circ}$ \cite{RakhmanThesis}. The corresponding final polarization state ($DOCP$, $\theta$) at the cavity exit was measured at the analysis station (Fig. \ref{fig:OpticsTable}) by using a rotatable quarter-wave plate (QWP$_2$), a Wollaston prism and two photo detectors (S$_1$ and S$_2$) that are mounted on integrating spheres.  A total of 18 data sets were used in a ROOT \cite{RootCernWeb} script to extract the transfer function parameters for mirrors M$_{s}$ and M$_{3}$. Fig. \ref{fig:TFCounterview} shows contour maps of the transfer functions for the left and right circularly polarized states obtained from these measurements. We have also used an auxiliary set of data measured for the circular polarization states of $\pm$97\% with $\theta$ being 10$^{\circ}$ -- 170$^{\circ}$ to validate the transfer function. The uncertainty in validation of the transfer function is,
\begin{equation}\label{eq:Uncrt}
\frac{\Delta DOCP_{\text{CIP}}}{DOCP_{\text{CIP}}} = \Bigg|\frac{DOCP^{m}_{\text{CIP}} - DOCP^{c}_{\text{CIP}}}{DOCP^{c}_{\text{CIP}}}\Bigg|
\end{equation}
where $DOCP^{m}_{\text{CIP}}$ and $DOCP^{c}_{\text{CIP}}$ are the measured and calculated values of $DOCP$ at the CIP respectively. The calculation shows that using $\pm$97\% data gives an average uncertainty of $\Delta DOCP_{\text{CIP}}$/$DOCP_{\text{CIP}}$ = 0.12\% in determination of $DOCP$ for a series of polarization states of $DOCP$ = $\pm$97\%.\\
\indent With the parameters extracted and validated with good precision ($<$ 0.5\%), the transfer function is fully established and can be used to determine the CIP polarization from any exit polarization state for a running cavity.
\subsection{Determination of intra-cavity polarization}
\begin{table*}[t!]
\caption{A comparison between the measured values of $DOCP$ and $\theta$ at the CIP without cavity mirrors and those calculated from the exit line measurement with cavity mirrors are in place. The measurement corresponds to the case when the QWP$_1$ angles were set at 50$^\circ$/-50$^\circ$ giving a nominal left/right polarization state at the CIP.}
\label{Table: CIPdocp}
\begin{center}
\begin{tabular}{c c |}\hline
  \multicolumn{2}{c|}{CIP (Measured)} \\\hline
$DOCP$ (\%)~~~~&~~~~$\theta$ (deg) \\\hline
 99.57 & 58.60 \\
-98.07 & 19.35  \\\hline
\end{tabular}
\hspace{-0.205cm}
\begin{tabular}{ c  c |}\hline
  \multicolumn{2}{c|}{CIP (Calculated)} \\\hline
$DOCP$ (\%)~~~~&~~~~$\theta$ (deg) \\\hline
99.26 & 83.52 \\
 -97.59 & 17.50  \\\hline
\end{tabular}
\hspace{-0.205cm}
\begin{tabular}{ c  c }\hline
  \multicolumn{2}{c}{Exit (Measured)} \\\hline
$DOCP$ (\%)~~~~&~~~~$\theta$ (deg) \\\hline
95.93   &    -61.05 \\
-97.32     & 16.21   \\\hline
\end{tabular}
\end{center}
\end{table*}
As shown in Eq. (\ref{eq:CmptAsym}), the precision in the measurement of $P_e$ depends on the accuracy of $P_{\gamma}$. Several factors can affect the accuracy of $P_{\gamma}$, such as the transfer function, the stability of intra-cavity polarization, the beam alignment and the birefringence of the optical elements.\\
\begin{figure*}[b!]
    \centerline{\includegraphics[width=7.8cm]{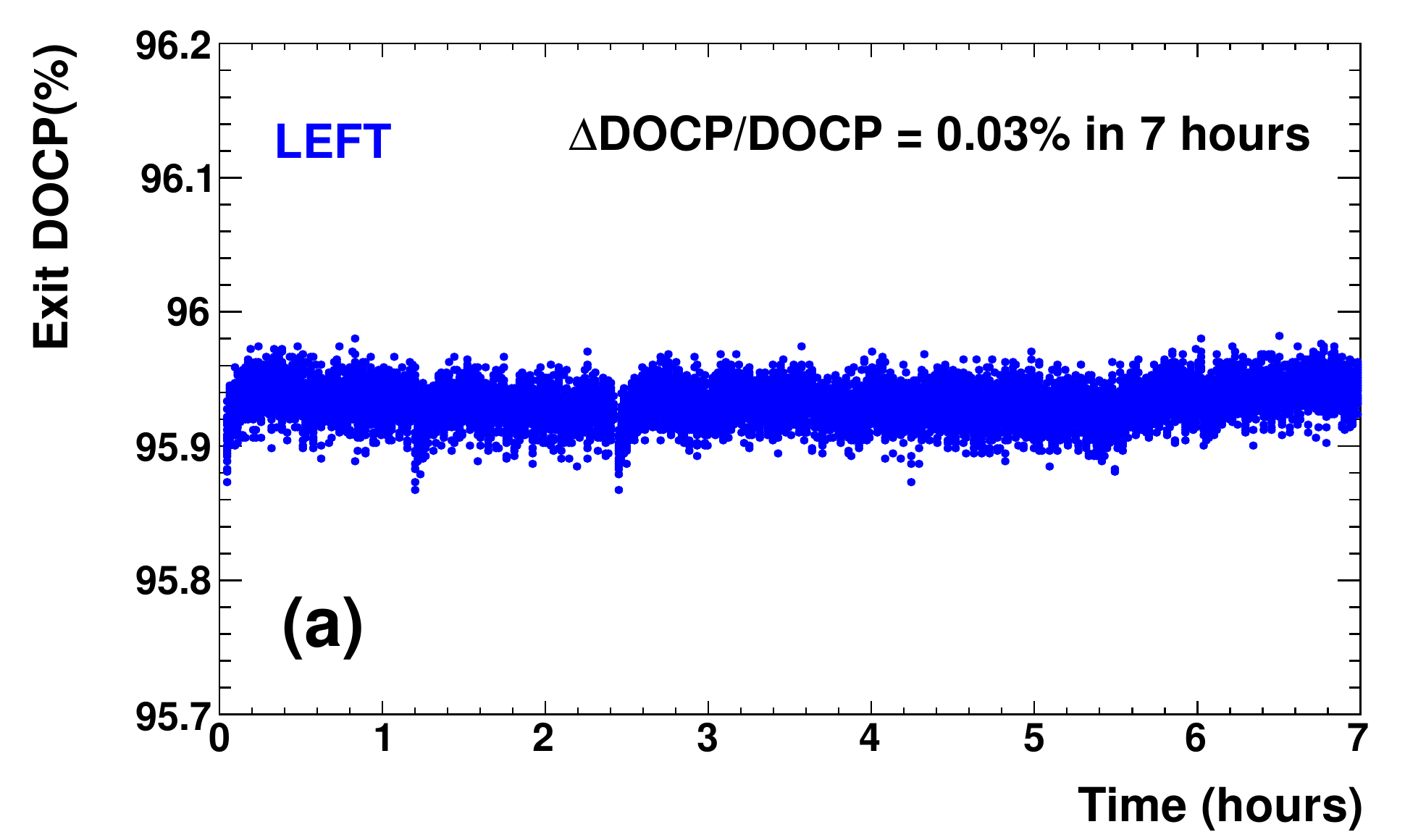}
    \hspace{0.5cm}
    \includegraphics[width=7.8cm]{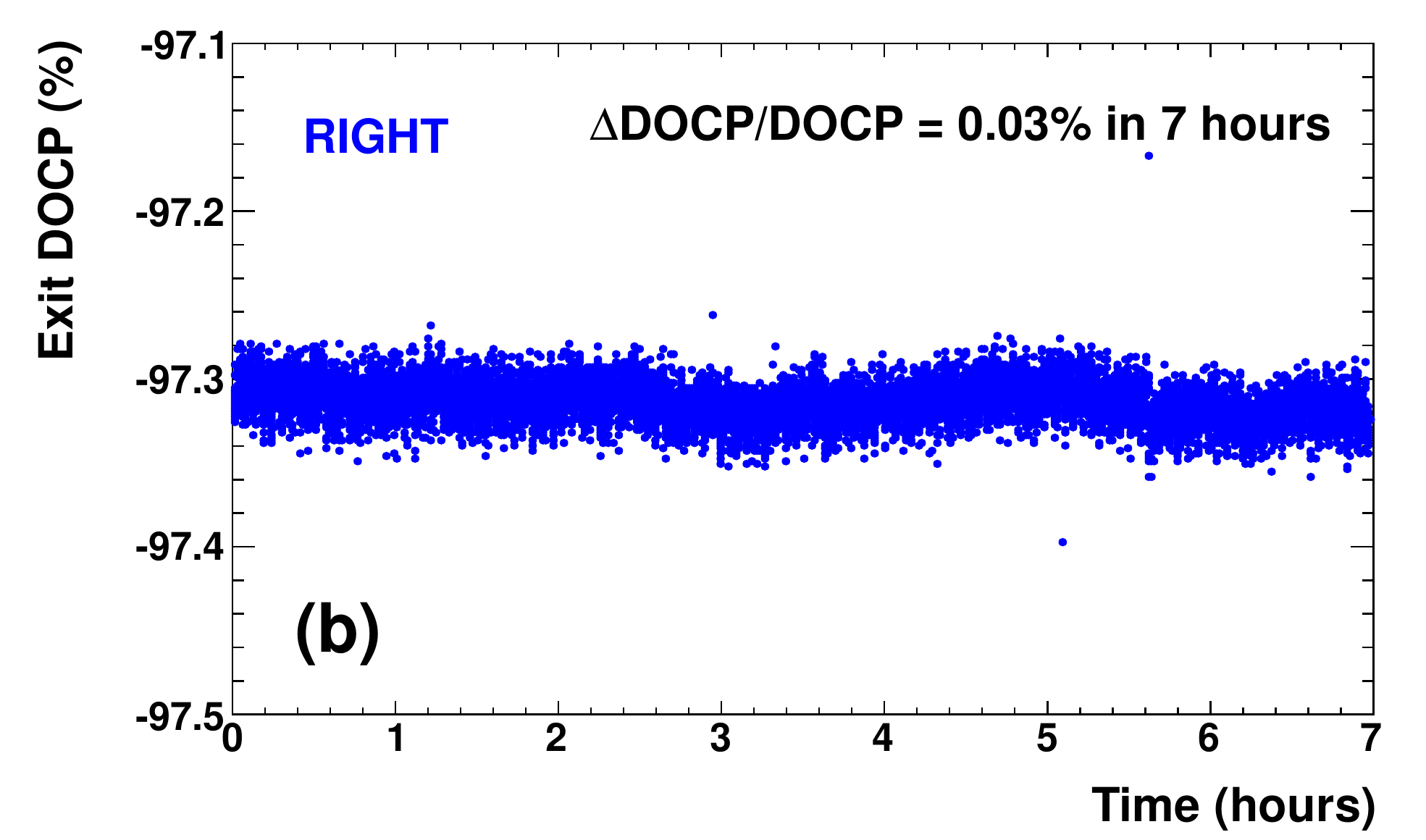}}
    \caption{The evolution of $DOCP$ for the left- (a) and right-circularly (b) polarized state at the cavity exit was monitored over 7 hours during the PREx experiment.}
    \label{fig:PolMont}
\end{figure*}
\indent Table \ref{Table: CIPdocp} shows a comparison between the measured values of $DOCP$ and $\theta$ at the CIP without cavity mirrors and the values predicted from the exit line measurement with cavity mirrors in place. The measurements correspond to the case when QWP$_1$ angles were set at 50$^\circ$/-50$^\circ$, giving nominal left/right polarization states at the CIP. The incoming beam alignment to the cavity was kept the same during operation of the cavity as when the transfer function was measured. We have validated the transfer function for the nominal polarization state at the CIP (99.57\%, -98.07\% in Table \ref{Table: CIPdocp}) created by QWP$_1$ by comparing the measured and calculated values of $DOCP$ at the CIP. The largest uncertainty in the calculation of the nominal $DOCP$ from the transfer function is calculated according to Eq. (\ref{eq:Uncrt}) and $\Delta DOCP_{\text{CIP}}$/$DOCP_{\text{CIP}}$ = 0.49\%. We believe that this uncertainty is mostly originating from various sources of birefringence in our system.\\
\indent Birefringence induced by multi-layer coatings of the cavity mirrors \cite{MicossiAPB, JacobOL, MoriwakiAPB} can be estimated according to the measured finesse. For our case, the maximum birefringence induced by multilayer cavity mirror coatings is estimated to be $\sim 10^{-4}$ rad for a measured finesse of $\sim$ 13,000. This effect causes~$\pm$0.1\% change between the two polarization states in $DOCP$ inside the cavity. The birefringence coming from the cavity mirror substrate was studied by measuring the difference in polarization of the signal transmitted through an uncoated substrate, and no degradation of polarization was observed.\\
\indent We have measured the $DOCP$ of the incoming laser beam before and after the vacuum window (W$_{e}$ and W$_{s}$ in Fig. \ref{fig:CavityFunctionalView}) without vacuum and the difference was negligible. However, when the cavity is under vacuum (10$^{-9}$ Torr), there could be a birefringence coming from the pressure difference between air and vacuum \cite{LoganOC}. Since the measurement of the transfer function can only be accomplished without cavity mirrors or vacuum it was not possible to evaluate this effect.\\
\indent The stability of the intra-cavity polarization is crucial to achieve precision Compton polarimetry. A thermoelastic deformation of cavity mirrors due to high power circulating inside the cavity could cause birefringence \cite{WinklerOC}. While the cavity is locked, we monitored the long-term stability of the exit line polarization online. Fig. \ref{fig:PolMont} shows a measurement for left- and right-circular polarization states monitored over 7 hours. The data indicates that the variations are at the order of 0.03\%.\\
\indent Other sources of systematic errors include the measurement of exit line polarization ($DOCP$ and $\theta$) and beam alignment by steering mirrors (M$_{1}$ and M$_{2}$ in Fig. \ref{fig:CavityFunctionalView}) to maximize the light coupling to the cavity is listed in Table \ref{Table: SysErr}.
\begin{table}[t!]
\caption{Summary of systematic errors on the measurement of the laser polarization at the CIP during the PREx experiment.}\label{Table: SysErr}
\begin{center}
\begin{tabular}{ l c }    \hline
Source of Error  ~~~~~~&  (\%)\\\hline
Validation of transfer function  ~~~~~~& 0.49\\
Transmitting through M$_{ce}$  ~~~~~~& 0.10\\
Transmitting through M$_{cs}$  ~~~~~~& 0.10\\
Variation in time  ~~~~~~& 0.03\\       
Variation of DOCP at exit line  ~~~~~~& 0.02\\
Variation of $\theta$ at exit line  ~~~~~~& 0.13\\      
Beam alignment to the cavity ~~~~~~& 0.10\\\hline
Total  ~~~~~~& 0.70\\\hline
\end{tabular}
\end{center}
\end{table}
\indent After taking into account all the systematic errors described, the total systematic error in the determination of the intra-cavity laser polarization was estimated to be 0.70\%. The averaged statistical errors during three-month period of the PREx experiment were 0.1\% and 0.13\% for the left- and right-circularly polarized states respectively. Based on the transfer function measurement, the average left- and right-laser polarizations ($P_{\gamma}$ = $DOCP_{\text{CIP}}$) with uncertainties during the PREx experiment are summarized as $P_{\gamma}^{L} = 99.26\% \pm 0.70\%~ \textnormal{(sys)} \pm 0.10\%~ \textnormal{(stat)}$, $P_{\gamma}^{R} = -97.59\% \pm 0.70\%~ \textnormal{(sys)} \pm 0.13\%~ \textnormal{(stat)}$.\\
\indent So far we have discussed how to make a reliable photon target with a pure circular polarization inside a high-finesse optical cavity for the electron beam so that there is efficient Compton scattering. The green cavity has been combined with the photon arm \cite{MeganNIMA} of the upgraded Compton polarimeter to continuously measure the electron beam polarization.
\section{Electron beam polarization measurement}
The electron beam polarization measurement can be achieved by one of the three ways at the down stream of the optical cavity: detecting the backscattered Compton photons, detecting the scattered Compton electrons, and detecting both the Compton photons and Compton electrons simultaneously. At the time we commissioned our laser system and cavity, the electron detector had not been fully commissioned and there is no way to detect the scattered Compton electrons. Therefore, the results presented here are solely based on the data from the upgraded photon detector and DAQ that are described in Ref. \cite{MeganNIMA}. In the following section, without going into the details, we will briefly present the results of the electron beam polarization measurement conducted for the first time during the PREx experiment with electron beam parameters of 1.06~GeV and 50~$\mu$A \cite{PRExPRL}.
\subsection{Compton spectrum}
\begin{figure}[b!]
    \centerline{\includegraphics[width=7.5cm]{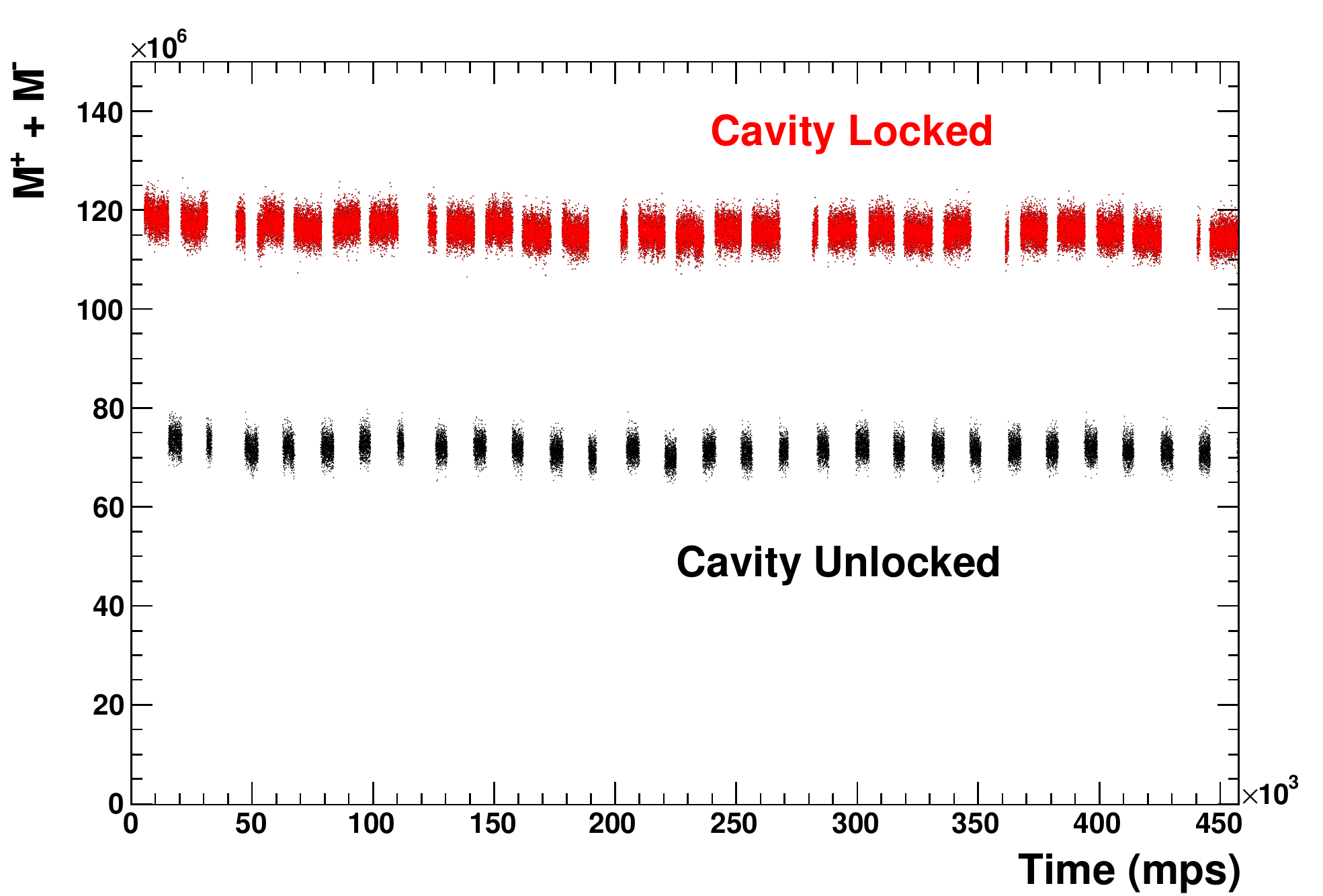}}
    \caption{Scattered Compton photon rates (red) along with the background rates (black), where each point is measured at an accelerator helicity timing signal (called the MPS signal) for a given accelerator helicity setting during a run.}
    \label{fig:TriggerRate}
\end{figure}
The upgraded Compton polarimeter measures the electron beam polarization based on the detection of scattered Compton photons during electron helicity reversal at a rate of 120 Hz \cite{SinclairPRSTAB}. The helicity of the electron beam at JLab can be flipped at a rate of between 1 ms to 33 ms at the injector with a pseudo-random pattern. Fig. \ref{fig:TriggerRate} shows the scattered photon rates versus accelerator helicity timing signal (called the MPS signal) measured with the photon detector for an electron beam energy and current of 1.06~GeV and 50~$\mu$A. During the data taking, the laser polarization is flipped periodically between right- and left-circular state and the cavity is locked for $\sim$ 90 seconds in each polarization state. Between each polarization state, the cavity is out of resonance for $\sim$ 40 seconds to measure the background rate, which is subtracted from the signal rate measured when the cavity is locked. Lower background rates can be achieved by several steps. Low vacuum pressure in the cavity (10$^{-9}$ Torr in our case) can reduce the scattering between electron beam and residual gas. Properly controlling the electron beam parameters upstream and carefully centering and focusing the beam at the interaction point is a key to achieve low background rates. During background rate optimization, the electron beam is scanned vertically until it crosses the photon beam at the CIP by using all four dipoles connected in series to the power supplies in the Compton chicane (Fig. \ref{fig:Chicane}).\\ 
\begin{figure}[t!]
    \centerline{\includegraphics[width=7.5cm]{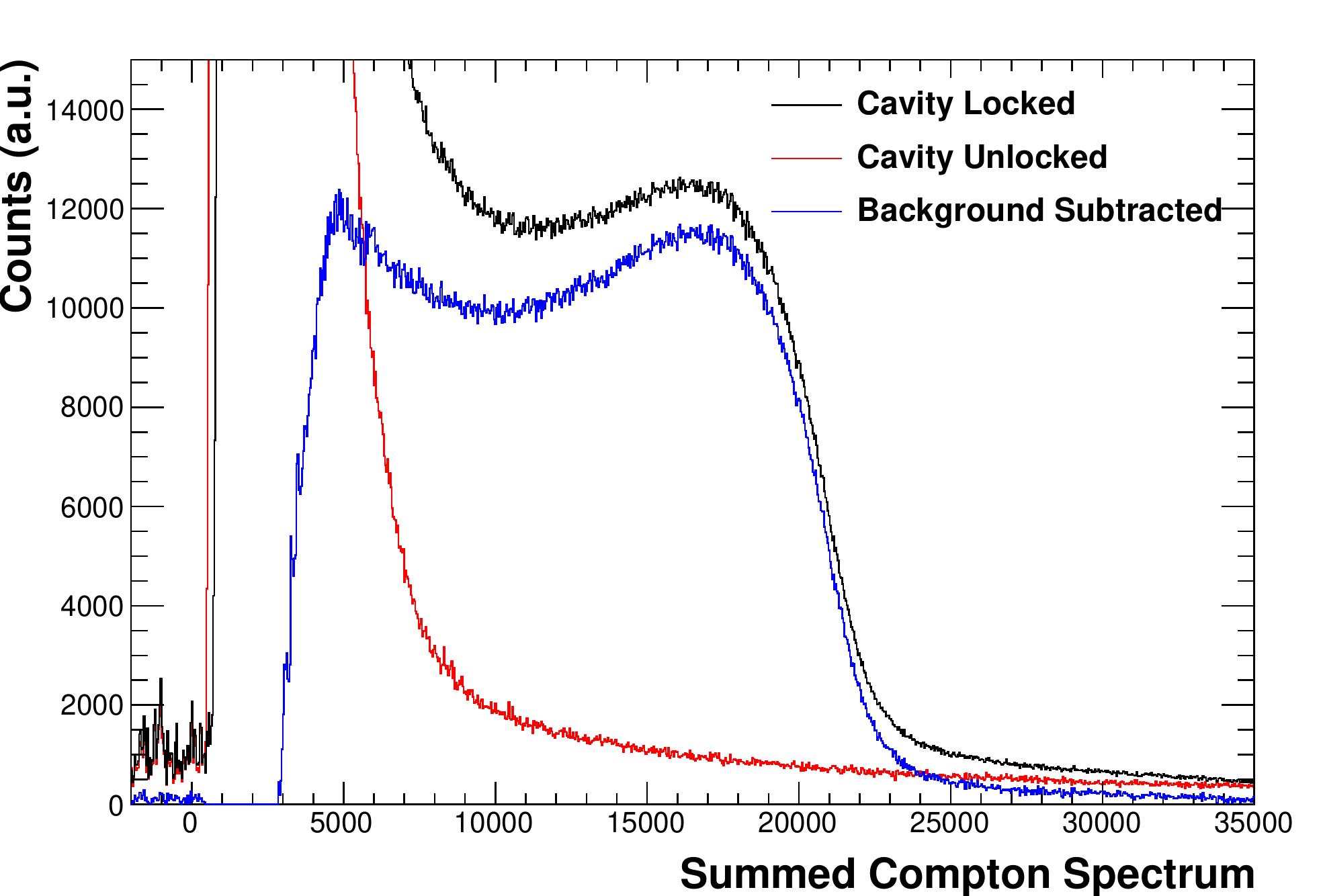}}
    \caption{Measured Compton photon energy spectra correspond to cavity locked (black), cavity unlocked (red) and background-subtracted (blue) cases. The first two cases (black and red) are out of the vertical scale in order to show the background-subtracted Compton spectrum (blue).}
    \label{fig:CptSpectr}
\end{figure}
\indent The Compton photon energy spectrum is measured using the triggered mode of the DAQ. In this readout mode, a discriminator with a very low threshold is used to trigger a readout of the FADC samples over a 500 ns window, allowing study of the sampled pulse shape and size. The sum of samples in the pulse represents the total energy deposited in the GSO crystal for each detected photon. This energy spectrum is plotted in (Fig. \ref{fig:CptSpectr}), with the horizontal axis in raw ADC units. To determine the experimental Compton asymmetry $A_{\text{exp}}$, the measured ADC response is calibrated to the well known Compton scattering photon energy spectrum and the result is compared to a full Monte Carlo simulation of the photon detector by GEANT4 \cite{Geant4}. After confirming good agreement with data, this same simulation is then used to calculate the theoretical asymmetry $A_{\text{th}}$ in Eq. (\ref{eq:CmptAsym}).
\subsection{Compton asymmetry}
\begin{figure}[t!]
    \centerline{\includegraphics[width=7.5cm]{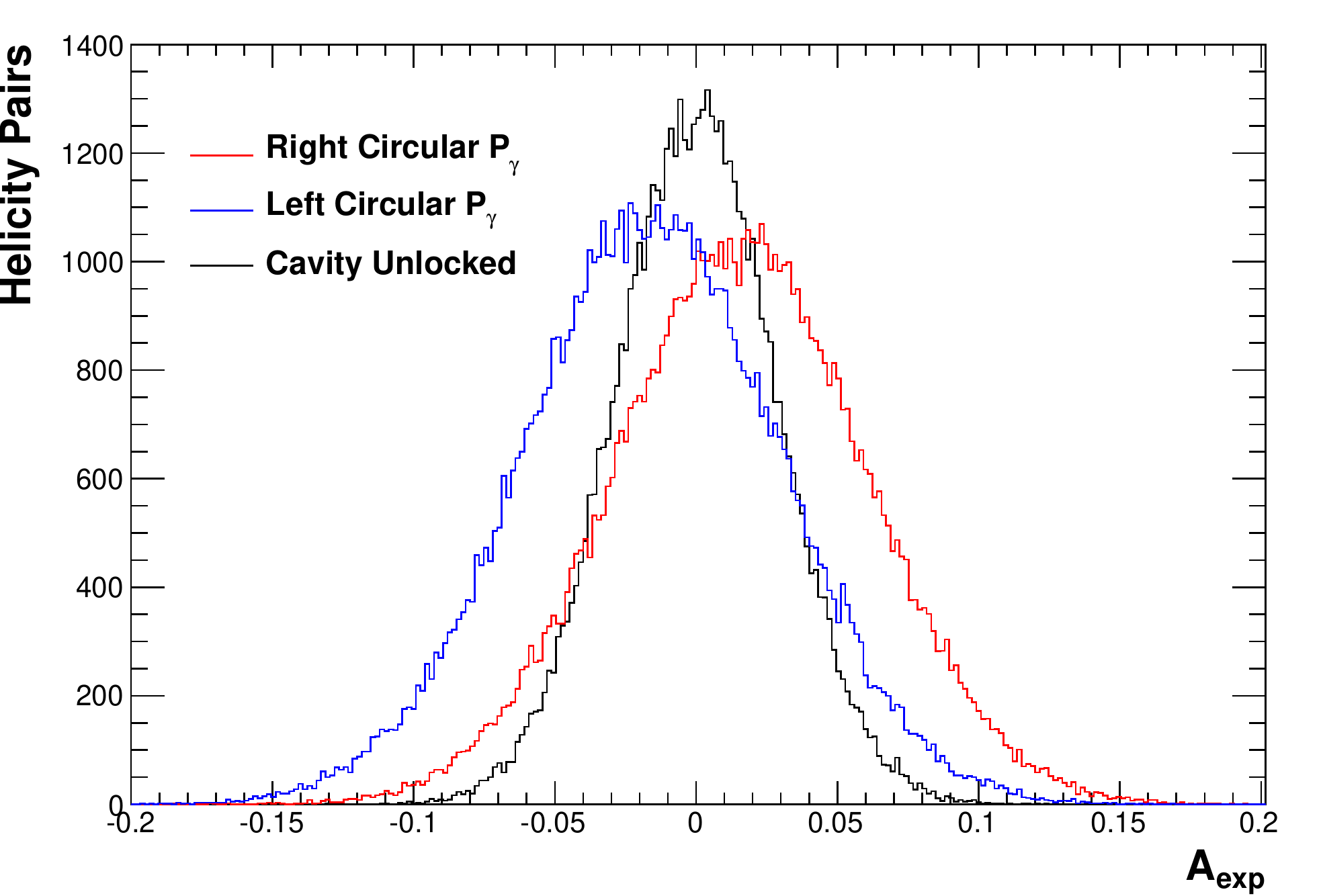}}
    \caption{Histogram of a background-subtracted Compton asymmetry taken for every electron helicity pair in a single one hour run. Colors represent right circular (red), left circular (blue) and cavity unlocked (black) cases.}
    \label{fig:Aexp}
\end{figure}
The integrating photon detector DAQ (as described in Ref. \cite{MeganNIMA}) sums the FADC samples over the entire period of stable beam helicity between flips, using a gate supplied by an accelerator timing board. These integrated signals ($S$) are used to calculate the asymmetry $A_{\text{exp}}$ as:
\begin{equation}\label{eq:AsymMegan}
A_{\text{exp}} = \frac{S^{+} - S^{-}}{S^{+}+ S^{-}},
\end{equation}
for each laser helicity period with separate sums of integrals for all electron helicity ($\pm$) windows. A separate sum and difference of integrated signals for the adjacent cavity-unlocked periods, $B = (S^{+}-S^{-})_{\text{unlocked}}/(S^{+} + S^{-})_{\text{unlocked}}$, used to determine background for the cavity-locked period. After taking into account the background and helicity-dependent electron beam parameters at the accelerator source, Eq. (\ref{eq:AsymMegan}) becomes,
\begin{equation}
A_{\text{exp}} = \frac{\langle M^{+} \rangle - \langle M^{-} \rangle}{\langle M^{+} \rangle + \langle M^{-} \rangle - 2\langle B \rangle},
\end{equation}
where $\langle ~ \rangle$ denotes the mean integral per helicity window over each cavity (locked/unlocked) period. Here, $M^{\pm}$ is the measured integrated signal with the laser locked, and includes both the Compton-scattered signal $S^{\pm}$ and the background signal $B$. $A_{\text{exp}}$ can be extracted separately for each laser polarization using one of the three averaging techniques discussed in Ref. \cite{MeganNIMA}. Fig. \ref{fig:Aexp} shows a histogram of a background-subtracted Compton asymmetry taken for every electron helicity pair in a single one hour run.
\subsection{Electron beam polarization}
\begin{figure}[t!]
    \centerline{\includegraphics[width=7.5cm]{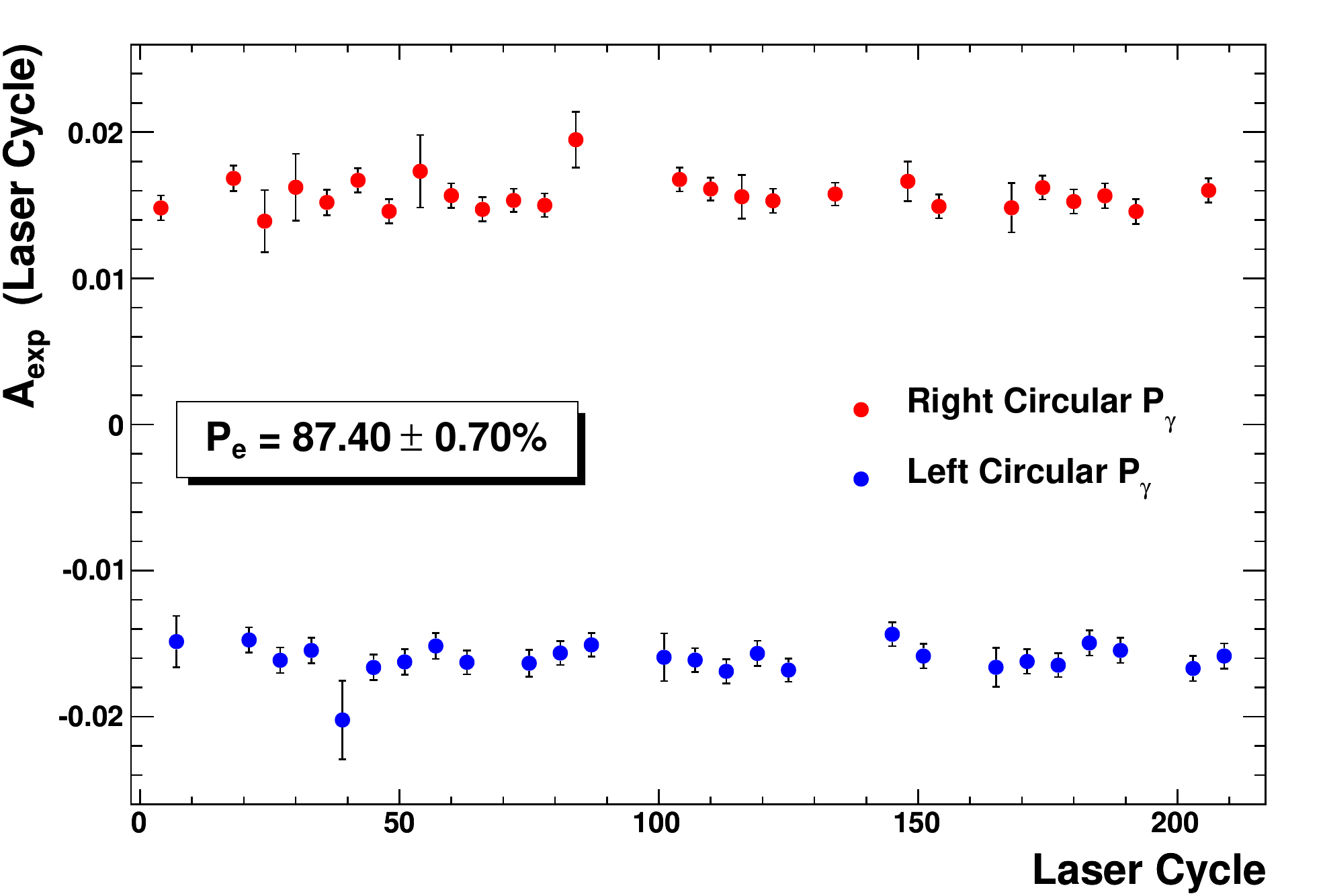}}
    \caption{Asymmetry versus left- (blue) and right-circularly (red) polarized laser cycles for an entire run. An average asymmetry is used for calculating the electron beam polarization for a typical run.}
    \label{fig:LaserCycle}
\end{figure}
The electron beam polarization is continuously monitored by the Compton polarimeter during the PREx experiment. As we described in Section \ref{sec:intro}, the electron beam polarization $P_{\text{e}}$ is extracted from the measured experimental asymmetry $A_{\text{exp}}$, the laser polarization $P_{\gamma}$ and the theoretical asymmetry $A_{\text{th}}$, according to Eq. (\ref{eq:CmptAsym}). Fig. \ref{fig:LaserCycle} shows the electron beam polarization measurement based on an average of two experimental asymmetry numbers obtained for the left and right laser cycles during a typical run. The average experimental asymmetry corresponding to the left and right laser cycles for this run are -1.59\% $\pm$~0.02\%~\textnormal{(stat)} and 1.56\% $\pm$~0.02\%~\textnormal{(stat)}. After taking the value of $A_{\text{th}}$ as 1.825\% and using the asymmetry numbers above, the extracted average electron beam polarization for this run is 87.40\% $\pm$~0.70\%~\textnormal{(stat)}.\\
\indent Finally, after taking into account all the uncertainties in other parameters (such as laser polarization, theoretical asymmetry, photon detector gain shift etc.), the average beam polarization measured by the Compton polarimeter during the PREx experiment is $P_{\text{e}} = 88.20\% \pm 0.12\%~ \textnormal{(stat)} \pm 1.04\%~ \textnormal{(sys)}$. It was compared with an independent M\o ller polarimeter measurement made at different times during the run which gave an average $P_{\text{e}} = 90.3\% \pm 0.1\%~ \textnormal{(stat)} \pm 1.1\%~ \textnormal{(sys)}$ \cite{PRExPRL}. The observed difference in electron beam polarization is consistent within the systematic errors between the two measurements.
\section{Conclusions}
The green Fabry-Perot cavity provided Hall A of JLab with a unique laser source to carry out precision Compton polarimetry. After combining with the upgraded Compton DAQ and photon detector \cite{MeganNIMA}, it extends the operating energy range of the previous Compton polarimeter \cite{Escoffier} from 3 -- 11~GeV to 1 -- 11~GeV. High-precision electron beam polarimetry was tested for the first time at JLab with beam energy and current of 1.06~GeV and 50~$\mu$A, and achieved its 1.0\% precision goal for the PREx experiment. The system we developed allows us to obtain a reliable intra-cavity power and stable photon beam polarization for the long-term operation of the upgraded Compton polarimeter. The cavity mechanical and locking stability was excellent during the three-month period of the PREx experiment, despite the concern of a very high-radiation and acoustically noisy environment in Hall A at JLab. The frequency-doubled green laser power was stable indicating that the PPLN crystal could withstand high power for an extended period of time.\\
\indent The frequency locking of the green laser generated by seeding the Nd:YAG NPRO laser to the YDFA makes the intra-cavity power scalable. This allows a possibility of boosting the intra-cavity power further (10~kW or more at 532~nm) by increasing the injection power to the cavity provided there is no thermal problem in the cavity mirrors. A higher intra-cavity power may also be achievable by replacing the current cavity mirrors with higher finesse mirrors which is possible with the current feedback electronics. Regarding the intra-cavity laser polarization, it is one of the dominant errors in our polarimeter with an uncertainty of 0.7\%. It is dominated by our ability to through direct measurement, effects that might alter the polarization over the numerous re-circulations within the cavity. Other factors include analyzing power and photon detector gain shift can also affect the systematic errors in the final electron beam polarization \cite{MeganNIMA}. Recent studies at HERA \cite{HERA_JIns} and JLab Hall C \cite{HallCCompton} show that it is possible to get $<$ 0.3\% level precision for the intra-cavity laser polarization with proposed techniques. So far this result is only based on scattered photon analysis. This measurement could be combined with an independent electron analysis as demonstrated in Hall C at JLab with a reported precision of 0.6\% \cite{HallCCompton}. All of this encourages us that, with a careful and dedicated study, it should be feasible to get $<$ 0.5\% precision at electron beam energies below 2~GeV.\\
\indent Future high-precision parity-violating electron scattering experiments at JLab, such as the MOLLER \cite {MOLLER}, PVDIS \cite{PVDIS}, PREx-II \cite{PREx2}, and CREx \cite{CREx} experiments, also require electron beam polarimetry with sub-1\% precision and would rely on the green Compton polarimeter we built. This requires an improvement of the current precision by at least a factor of two. Finally, a Compton polarimeter with a mode-locked pulsed-laser-based optical cavity \cite{KEK1, KEK2} would give a much higher signal-to-noise ratio in scattered Compton events and would make the system even more robust and efficient, especially in the more noisy electron beam environment expected \cite{CmptPRSTAB} after the upgrade of the CEBAF energy to 12~GeV.
\section*{Acknowledgements}
The authors wish to thank the Hall A technical staff for their assistance in installing the upgraded system and the JLab Accelerator Division for their efforts to achieve and maintain the high electron beam quality necessary for running the Compton polarimeter. This work is supported by DOE grants DE-AC05-06OR23177, under which Jefferson Science Associates, LLC, operate Jefferson Lab, and DE-FG02-84ER40146.
%
%

\section*{References} 





\end{document}